\documentclass[vipsnames, 10pt]{article}
\usepackage{titling}
\newcommand{\subtitle}[1]{%
  \posttitle{%
    \par\end{center}
    \begin{center}\LARGE#1\end{center}
    \vskip0.5em}%
}
\usepackage{array}

\usepackage{amsmath,amsfonts,amsthm,bm,mathtools} 
\usepackage{caption} 				
\usepackage{subcaption} 				
\usepackage[utf8]{inputenc}
\usepackage[english]{babel}
\usepackage{blindtext}
\usepackage{geometry}
\usepackage{array}
\usepackage{multirow}
\usepackage{tabu}
\usepackage[utf8]{inputenc}
\usepackage{xargs}                      		
\usepackage[pdftex,dvipsnames]{xcolor} 	
\usepackage[colorinlistoftodos,prependcaption,textsize=tiny]{todonotes} 
\usepackage{graphicx}
\usepackage{afterpage}
\usepackage[font=small,labelfont=bf,
   justification=justified,
   format=plain]{caption} 
\usepackage[colorlinks=blue]{hyperref}
\hypersetup{
    colorlinks=false,
    pdfborder={0 0 0},
}
\hypersetup{
    colorlinks=true,
    linkcolor=blue,
    citecolor= blue,
    filecolor=blue,      
    urlcolor=blue,
}
\usepackage[square,sort,comma,numbers]{natbib}
\usepackage{sectsty}
\sectionfont{\centering}

\usepackage[ bf]{titlesec}
\usepackage{algorithm}
\usepackage[noend]{algpseudocode}
\usepackage{blindtext,titlefoot}
\title{\vspace{-0.5cm}
{ 
\Large \bf
{A Look into Chaos Detection through Topological Data Analysis}%
}\\
}

\author{
{
Joshua R.~Tempelman 
and 
Firas A.~Khasawneh
}
}

\date{\small
{\it
 Department of Mechanical Engineering, Michigan State University, 428 S. Shaw Ln. East Lansing, MI 48824}\\
}\vspace{-1.75cm}
\usepackage{fancyhdr}
\usepackage[T1]{fontenc}

\usepackage{multicol}
\usepackage{xcolor}
\definecolor{black}{rgb}{0,0,0}

\usepackage{titling}

\begin{document}


\unmarkedfntext{\noindent \hspace{-16pt} 
	{\small Address correspondence to \href{mailto:khasawn3@egr.msu.edu}{\textit{khasawn3@egr.msu.edu}}} 
}


\unmarkedfntext{\noindent \hspace{-16pt} 
	{\small
		Code and data available at Mendeley repository \href{https://data.mendeley.com/datasets/4kszknf6vj/2}{
			\textit{Chaos Detection with Persistent Homology}
		} 
}}

\unmarkedfntext{\noindent \hspace{-16pt} 
	{\small
	{Accepted February 26,~2020 in \href{https://www.journals.elsevier.com/physica-d-nonlinear-phenomena}{\textit{Physica D: Nonlinear Phenomenon}}, 
	{\fontfamily{lmss}\selectfont{DOI}}: \href{https://www.sciencedirect.com/science/article/pii/S0167278919300739?via\%3Dihub}{\textit{10.1016/j.physd.2020.132446}}
}
}
}
\maketitle

\titlespacing\section{0pt}{16pt plus 4pt minus 2pt}{02pt plus 2pt minus 2pt}
\titlespacing\subsection{0pt}{12pt plus 4pt minus 2pt}{2pt plus 2pt minus 2pt}
\titlespacing\subsubsection{0pt}{12pt plus 4pt minus 2pt}{2pt plus 2pt minus 2pt}

\vspace{-.5cm}

\vspace{-.75cm}

\renewcommand{\abstractname}{}    

\begin{abstract} 
\noindent
\rule{5.6 in}{1pt}
\\
\noindent
\textbf {Abstract.}
Traditionally, computation of Lyapunov exponents has been the marque method for identifying chaos in a time series. Recently, new methods have emerged for systems with both known and unknown models to produce a definitive 0--1 diagnostic. However, there still lacks a method which can reliably perform an evaluation for noisy time series with no known model. 
In this paper, we present a new chaos detection method which utilizes tools from topological data analysis. 
Bi-variate density estimates of the randomly projected time series in the  $p$-$q$ plane described in Gottwald and Melbourne's approach for 0--1 detection are used to generate a gray-scale image.
We show that simple statistical summaries of the 0D sub-level set persistence of the images can elucidate whether or not the underlying time series is chaotic.
Case studies on the Lorenz and Rossler attractors as well as the Logistic Map are used to validate this claim.
We demonstrate that our test is comparable to the 0--1 correlation test for clean time series and that it is able to distinguish between periodic and chaotic dynamics even at high noise-levels. 
However, we show that neither our persistence based test nor the 0--1 test converge for trajectories with partially predicable chaos, i.e. trajectories with a cross-distance scaling exponent of zero and a non-zero cross correlation. 
  \\

\noindent
\textbf {Key words. }%
Topological data analysis, chaos detection, persistent homology, time series

\noindent
\rule{5.6 in}{1pt}
\end{abstract}

\vspace{-.35cm}

\section{Introduction}
Chaos is an obscure phenomenon which appears in many physical and theoretical systems. 
Chaotic systems display a hyper-sensitivity to initial conditions, and their subsequent dynamics are difficult to anticipate as they are seemingly random and sporadic. 
Differentiating a chaotic time series from a periodic one is not a trivial task. 
The traditional recipe for doing so utilizes a tangent space method where, for example, the sign of the maximal Lyuponov exponent is used to classify the systems behavior. 
This method has proven reliable, although it is not straightforward to implement for systems with no known model, even with the procedures given by Bennetin~\cite{Benettin1980} and Wolf {\it{et al}}.~\cite{Wolf1985}. \textcolor{black}{ Additionally, the tangent space methods perform poorly when dealing with noisy finite time series, and this is a major limitation when analyzing data from a high-dimensional and noisy system.  }

\textcolor{black}{
	There have been recent efforts to detect chaos in time series such as the binary test introduced by  Wernecke {\it{et al}}.~\cite{Wernecke2017} which can also distinguish strong chaos from partially predictable chaos.
	Like the Lyuponov exponent procedure, this method is reliant on the time--evolution of initially close trajectories.
	Thus, a known model is required to deliver a diagnosis. 
	Additionally, there are no results given in~\cite{Wernecke2017} on the effects of noise on the scoring method. 
}

\textcolor{black}{
	A popular emergent tool for chaos detection in a time series is the 0--1 test which was introduced by Gottwald and Melbourne in 2004~\cite{Gottwald2004}. 
	Unlike tangent space methods, knowledge of the full state--space is not required to perform the 0--1 test. 
}
\textcolor{black}{
	Since a state-space reconstruction is not required, the 0--1 method is favorable for data streams for which not all state variables are  observable; this is the case for many engineered and natural systems. 
}

\textcolor{black}{
	The 0--1 test has been presented in two forms: the regression test and the correlation test (see Section~\ref{0--1test} and  Refs.~\cite{Gottwald2004,Gottwald2005,Gottwald2009,Skokos2016}). 
	The correlation test has been shown to perform exceptionally well for noise-free time series~\cite{Gottwald2009}. 
	However, when noise contaminates the data, a modified version of the original regression test desensitizes the scores to noise~\cite{Gottwald2005,Gottwald2009}. Alternatively, the correlation method handles noise exceptionally well if ``gauge'' values are given for a time series known to be periodic~\cite{Gottwald2016}. However, the required gauge score  is not always obtainable in real world systems. 
}

\textcolor{black}{
	In this paper, we present an alternative approach for combating the sensitivity of the 0--1 test to noise. Our approach combines elements from the 0--1 recipe with a tool from Topological Data Analysis (TDA) specifically persistent homology. 
	We leverage the stability of persistent homology in the presence of noise to deliver a method for identifying the shift-points between chaotic and regular dynamics in moderately noisy time series.
}
\textcolor{black}{
}

The paper is organized as follows. In Sections \ref{lorenz} and \ref{0--1test}, we introduce the dynamic regimes we are studying and review the 0--1 test, respectively. 
In Section \ref{a_tda_approach}, we describe persistent homology and how this tool can be used in conjunction with projections of the time series data in order to identify the shift point between chaotic and periodic dynamics.  
Section \ref{robustness_to_noise} investigates the robustness of our approach to noise, and the summarizing conclusion is given in Section~\ref{conclusion}.

\subsection{Definitions of the Studied Dynamic Behavior}
\label{lorenz}
\begin{figure}[h]
	\centering
	\includegraphics[width = \textwidth]{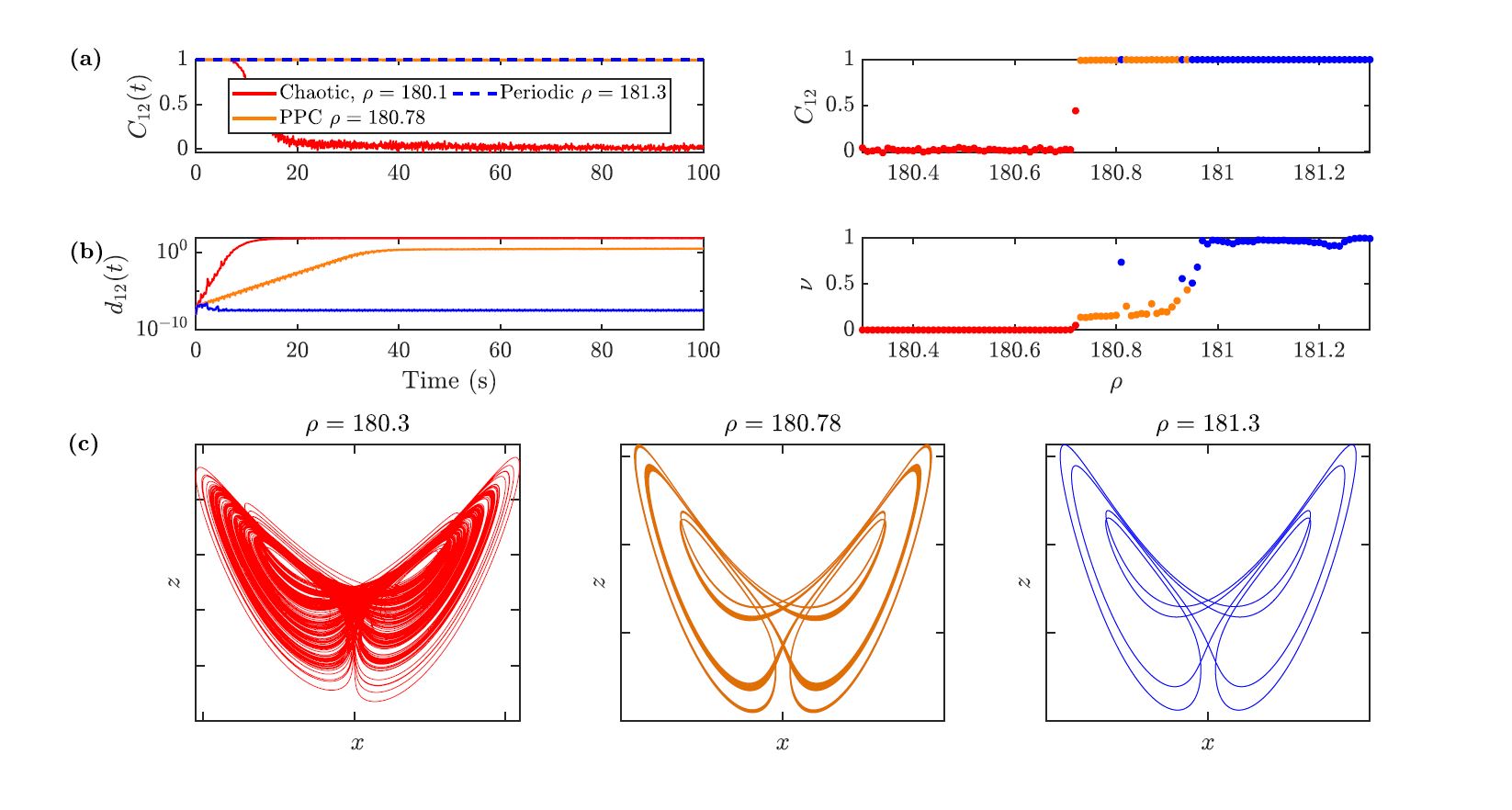}
	\caption{ (a) The cross-correlation $C_{12}$, (b) the scaling distance $\nu$, and (c) the various dynamic regimes displayed by the Lorenz system for modification of the parameter $\rho$. A chaotic trajectory is shown in for $\rho = 180.1$. Partially predictable chaos is shown in the center with the trajectory tightly orbiting the attracting braid, and fully periodic behavior is shown to the right where the trajectory is stable on the manifold. 
	}
	\label{fig:Lorenz_Wernecke}
\end{figure}

\textcolor{black}{
	The exposition of our method uses simulations of the Lorenz model.
	The equations of motion for this dynamical system are defined as~\cite{Lorenz1963}\\
	\begin{equation}
	\dot{x} = \sigma(y-x), \hspace{15pt} \dot{y} = x(\rho-z)-y, \hspace{15pt} \dot{z} = xy-\beta{z}.
	\end{equation}
	where nominal definitions of the Lorenz constants for this study are $\beta = 8/3$, $\sigma = 10$, and $\rho\in[180.3,\ 181.3]$ is left as a bifurcation parameter. 
	The manipulation of $\rho$ can lead to three distinct dynamic regimes as shown in Fig.~\ref{fig:Lorenz_Wernecke}.
	In~\cite{Wernecke2017}, it is shown that the Lorenz system displays periodic, partially predictable chaos (PPC), and chaotic behavior over the range $\rho\in[180.3,\ 181.3]$. 
	These regimes are formally defined in~\cite{Wernecke2017} as strong chaos, PPC, and laminar flow. 
	%
	We implement the 0--1 tests of~\cite{Gottwald2005,Gottwald2009} as well as our proposed persistence--based method on time series which were shown in~\cite{Wernecke2017} to be PPC. 
	The PPC trajectories are characterized by having a non-zero cross correlation for time-scales beyond the Lyapunov prediction time and a cross-distance scaling exponent $\nu$ of zero. 
	The cross-correlation is defined as
	\begin{equation} 
	C_{12} = 
	\frac{1}{s^2}
	\langle \left[{\bf{x}}_1(t) -E(\textbf{x}_1)\right]\cdot \left[({\bf{x}}_2(t) -E(\textbf{x}_2)\right]	\rangle,
	\end{equation}
	where $E(\textbf{x})$ is the expected value given by
	$$
	E({\bf x}) = \lim_{N\to\infty}\frac{1}{N}\sum_{j=1}^{N}{\bf x}(j),
	$$ $\langle\hspace{3pt}\rangle$ denotes the ensemble mean, and the variance $s^2$ is  
	\begin{equation} 
	s^2 = \lim_{T-\infty}\int_{T}^{2T}\left[{\bf{x}}(t) -E(\textbf{x})\right]^2{\rm d}t.
	\end{equation}
	The cross-distance scaling between trajectories $\textbf{x}_1$ and $\textbf{x}_2$ is defined with
	\begin{equation} 
	d_{12}(t>T_{\lambda})\propto \delta^\nu , \ \ \ \ \  d_{12}(t) = \langle \left|{\bf{x}}_1(t) - {\bf{x}}_2(t)\right|\rangle
	\end{equation}
	where $T_{\lambda}$ is the Lyapunov prediction time $T_{\lambda} = \ln\left(|\textbf{x}_1-\textbf{x}_2|/\delta\right)/\lambda_m$,
	$\delta$ is the distance of initially close trajectories, and $\lambda_m$ is the maximal Lyapunov exponent. 
	The reader is directed to~\cite{Wernecke2017} for a detailed explanation and demonstration of these definitions.
}

\textcolor{black}{
	In this paper, the regimes will be referred to as chaotic, PPC, and periodic dynamics, respectively. 
	The ensemble mean of $10^4$ independent trials used in~\cite{Wernecke2017} is very effective in identifying these respective regimes with a binary diagnostic (Fig.~\ref{fig:Lorenz_Wernecke}). 
	However, these evaluations require a known set of differential equations to generate neighboring initial conditions. Thus, the approach in~\cite{Wernecke2017} requires the full knowledge of the state--space and cannot be applied to a single observation of a high--dimensional system. 
}

\subsection{The Traditional 0--1 Test}
\label{0--1test}
\textcolor{black}{
	A 0--1 diagnostic for identifying chaos in time series with known or unknown models, i.e. using recorded data, is given in~\cite{Gottwald2004}. The method has been presented in several different forms since its original conception~\cite{Gottwald2004, Gottwald2005, Gottwald2009,Skokos2016}. In this paper, we use the modified regression method described in~\cite{Gottwald2005} as well as the correlation method as described in~\cite{Skokos2016}. 
	The original test identifies some observable characteristic $\phi(j)$ of a dynamic system and uses it to construct planar $(p,q)$ data sets according to
	\begin{equation}
	p(n) = \sum_{j=1}^{n}\phi(j)\cos{jc},  \hspace{15pt}  {\mathrm{ and }}   \hspace{15pt} q(n) = \sum_{j=1}^{n}\phi(j)\sin{jc},
	\label{eq:pq}
	\end{equation}
	where $c$ is a random variable drawn from a uniform distribution in $(0, \pi)$ and $n=\left(0,1,\dots,{N}\right)$. Using a single data set of length ${N}$, the number of $c$ values selected creates $N_c$ data sets in the $(p,q)$ plane. In this paper, $N_c=100$ and $N=5000$ unless otherwise stated. The $p$-$q$ data is then used to compute the mean-square displacement
	\begin{equation}
	M_c(n) = \lim_{N \to \infty}\frac{1}{N}\sum_{j=1}^{N}\left[p_c(j+n)-p_c(j)\right]^2 + \left[q_c(j+n)-q_c(j)\right]^2.
	\label{eq:MSD}
	\end{equation}
	\textcolor{black}{ 
		Note that in Eq.~\eqref{eq:MSD}, $n = \left(1,2\dots, n_{\rm cut}\right)$ where $n_{\rm cut} = N/10$. The test for chaos is based on the asymptotic growth of $M_c(n)$ with respect to $n$, which is a bounded function of time for periodic dynamics and grows linearly with time for chaotic dynamics~\cite{Gottwald2009}.  
		It is shown in~\cite{Gottwald2009} that the modified mean-square displacement $D_c(n)$ possesses better convergence characteristics. 
		To construct $D_c(n)$, begin by defining the oscillatory term
		\begin{equation}
		V_{\rm osc}(c,n) = \left(E\phi\right)^2\frac{1-\cos\left(nc\right)}{1-\cos\left(c\right)}.
		\end{equation}
		Now, subtract the $V_{\rm osc}$ term from the mean-square displacement to define the modified mean-square displacement
		\begin{equation}
		D_c(n) = M_c(n) - V_{\rm osc}.
		\end{equation}
		The test can now be performed using either the correlation test or regression test. Using the standard definitions for covariance and variance respectively
		\begin{equation}
		{\rm cov}\left(x,y\right) = \frac{1}{q}\sum_{j=1}^{q}\left(x\left(j\right)-\bar{x}\right)\left(y\left(j\right) - \bar{y}\right),
		\hspace{10pt}{\rm and}\hspace{10pt}
		{\rm var}\left(x\right) = {\rm cov}\left(x,x\right),
		\end{equation}
		the correlation $K_c$ value is found using the correlation function
		\begin{equation}
		K_c = {\rm corr}\left(\xi, \Delta\right) = \frac{{\rm cov}\left(\xi, \delta\right)}{\sqrt{ {\rm var}\left(\xi\right){\rm var}\left(\Delta\right)}}\in[-1,1],
		\label{eq:corr}
		\end{equation}
		where $\xi = \left(1,2,\dots,n_{\rm cut}\right)$ and  $\Delta = \left(D_c(1), D_c(2), \dots D_c\left(n_{\rm cut}\right)\right)$.
		The 0--1 correlation $K$ value of the time series is then defined as the median value of the $K_c$ values with a median near zero noting a periodic time series and a median near one noting a chaotic time series. }
}

\textcolor{black}{
	Alternatively, the 0--1 regression test may be applied according to 
	\begin{equation}
	K_c = \lim_{n\to\infty}\frac{\log{\tilde{D}_c(n)}}{\log{n}}, \ \ \ \tilde{D_c}(n) = D_c(n)-\min_{n=1,\dots,n_{\rm cut} }D_c(n).
	\end{equation}
	In~\cite{Gottwald2005}, a $K_c$ value is computed for $M_c(n)$ rather than $\tilde{D}_c(n)$, and this will be referred to as the modified 0--1 regression test
	\begin{equation}
	K_c = \lim_{n\to\infty}\frac{\log{M_c(n)}}{\log{n}}.
	\end{equation} 
	For finite data, the $K_c$ score is computed as the slope of the line which fits the $\log(M_c(n))$ versus $\log(n)$ with the least absolute deviation~\cite{Gottwald2009}.
	The modified 0--1 regression test will be used instead of the standard regression test in this paper since it has been shown to perform better in the presence of noise~\cite{Gottwald2005,Gottwald2009}; noise robustness will play major role in later sections of this paper (section~\ref{robustness_to_noise}).
}

The 0--1 test typically works better than tangent space methods for detecting chaos. However, some abating factors must be addressed. 
The test will always return a periodic diagnosis when the data is over-sampled for the case of time-continuous systems~\cite{Melosik2016}. 
Additionally, false positives are found when periodic time series are contaminated with substantial noise~\cite{Gottwald2005}.  
The issue of oversampling can be resolved by properly sub-sampling the time series according to~\cite{Gottwald2009,Melosik2016, Myers2019, Myers2019a}, the effects of noise are more difficult to overcome. 
A method has been developed which allows the 0--1 test to handle noise very effectively if a gauge-value (i.e. time series with known periodic parameters) is provided~\cite{Gottwald2016}.
Furthermore,  the effect of noise on the 0--1 test has been leveraged to predict the noise level within a periodic time series~\cite{Skokos2016}. 
\textcolor{black}{
	If noise is of concern and using a gauge value is not an option, the modified regression score should be used~\cite{Gottwald2005,Gottwald2016}.  
	However, whereas the 0-1 test as described above has a mathematically rigorous underpinning~\cite{Gottwald2009a}, this is not the case for  the more noise-robust modification~\cite{Gottwald2005}.
}

\begin{figure}[h]
	\centering
	\includegraphics[width = .75\textwidth]{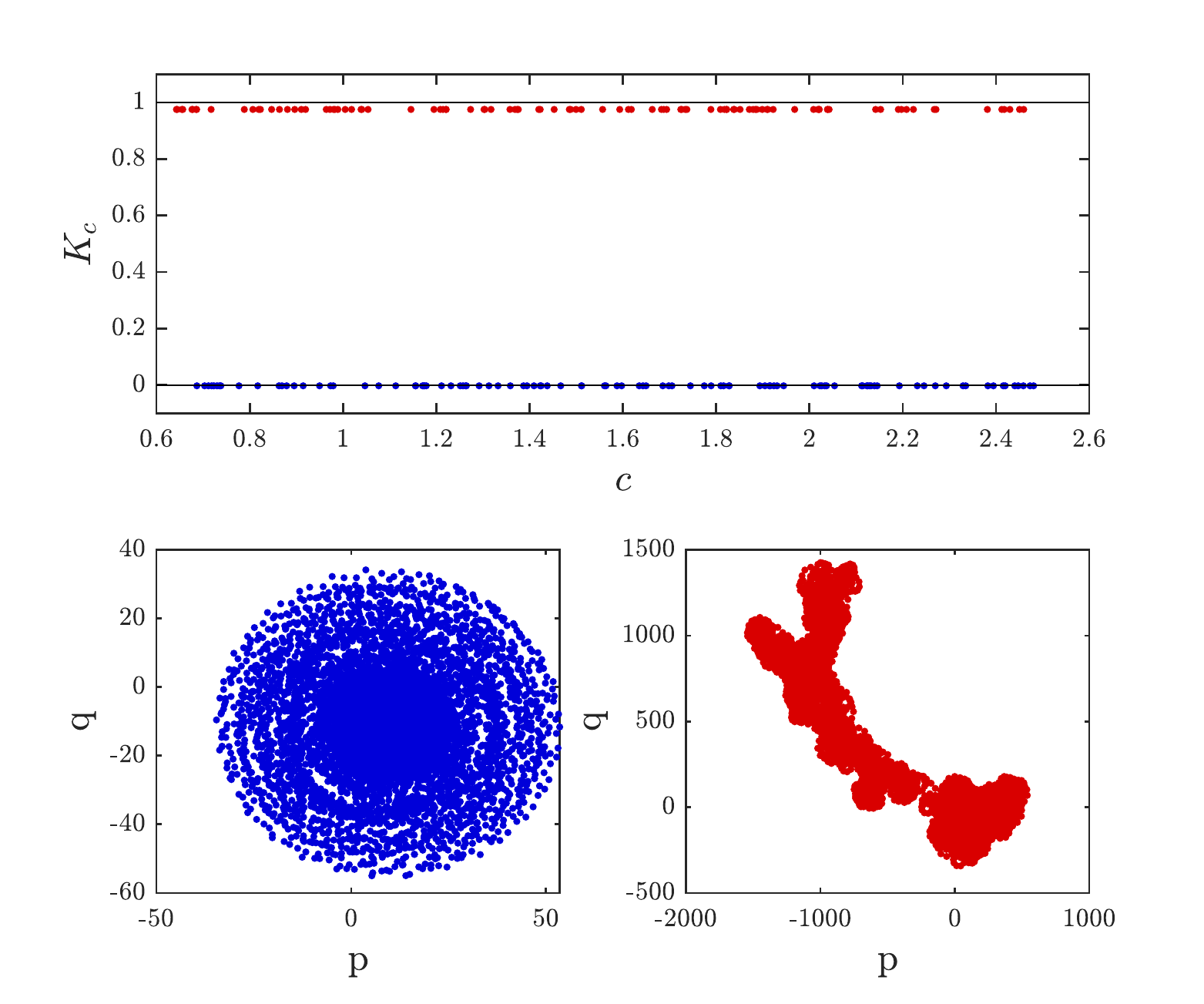}
	\caption{(top) the $K_c$ values found using the 0--1 method for a periodic time series (blue) and a chaotic time series (red). The data is generated with the Lorenz model. (Bottom left) The $p$-$q$ projections for the periodic time series produce a compact annular geometry. (Bottom right) The $p$-$q$ projection of the chaotic time series produces a scattered, irregular pattern.}
	\label{fig:p-q_projections}
\end{figure}

An interesting intermediate result of the 0--1 method is the projection of the data onto the $p$-$q$ plane. If the dynamics of the time series are periodic, then the resulting projection will have a compact and typically annular topology.
In contrast, Fig.~\ref{fig:p-q_projections} shows that chaotic dynamics lead to an irregular scattering of points in the $p$-$q$ space. Therefore, the topology of the $p$-$q$ projections can give insight into the underlying dynamics, as we show in Section \ref{a_tda_approach}.

\section{A Topological Approach}
\label{a_tda_approach}
The motivation behind a TDA approach is that data can have shape, and that shape has a meaning. TDA provides tools which lead to a computational summary of the shape of data. 
\textcolor{black}{
	We showed in Section~\ref{0--1test} that the 0--1 test produces data projections with meaningful topology. Thus, these projections can be quantified in a topological sense and studied in a new framework.
}

The ability to quantify the topology of a data structure has led to many breakthroughs in various fields. Specifically, TDA has already made meaningful contributions in time series analysis~\cite{Robinson2014,Perea2018, Khasawneh2018}, economics~\cite{Gidea2018}, machining dynamics~\cite{Khasawneh2016,Khasawneh2018a, Yesilli2019,Yesilli2019a}, biochemistry~\cite{Offroy2016}, and plant morphology~\cite{Li2018}. In this section, we extend the application of this powerful computational tool to chaos detection.

\textcolor{black}{
	Recently, Mittal and Gupta have explored persistence homology as a tool for detecting early bifurcations and chaos in dynamic systems~\cite{Mittal2017}. In~\cite{Mittal2017}, persistent homology is used to identify the presence of holes in the phase portraits of deterministic dynamical systems, thus an indicator for bifurcations. The method has shown promising results in~\cite{Mittal2017} even with a signal to noise ratio of 30 dB. However, wavelet filtering is required to pre--process noisy data.  Additionally, we note that their method requires one to have access to the whole phase-space or to perform a phase-space reconstruction. The approach we describe here requires neither data pre--processing nor phase-space reconstruction.   
}

We use the Lorenz system as a test model for elucidating our TDA-based approach for chaos detection. Time series which are generated for different values of the bifurcation parameter $\rho$ in the Lorenz model  are used to exhibit chaotic, PPC, and periodic dynamics. The Lorenz time series were produced using the ODE45 function in MATLAB\textsuperscript{\textregistered} with a relative tolerance of 1e-6, a time step of $\Delta{t}=0.001$, and a transient cut-off of 100 seconds.  

\textcolor{black}{
	Consistent with the 0--1 test, the time series data must first be sub-sampled, if needed. 
	This was done using the frequency approach presented by Melosik and Marszalek~\cite{Melosik2016, Myers2019}. 
	This sub-sampling method results in a time series sampled such that $2f_{\rm max}<f_s<4f_{\rm max}$ where $f_s$ is the sub--sampling frequency and $f_{\rm max}$ is the maximum significant frequency of the data in the frequency domain. 
	In the 0--1 test, failure to sub-sample may result in false-negatives.
	Similarly, we found that oversampling affects the topological characteristics of the $p$--$q$ projections thus leading to incorrect conclusions when studying the topology of the $p$--$q$ space.
	As a general rule-of-thumb, it is suggested that the sampling rate chosen in the time series sub--sampling is set to three times the maximum significant frequency of the data.   
}

\textcolor{black}{
	The subsequent sections give a detailed development of the TDA approach; the overview is as follows. Multiple projections of $p$-$q$ data are generated for a given time series. The bi-variate Kernel Density Estimates (KDEs) of these projections produce a gray scale image (Fig.~\ref{fig:method}). 
	Sub-level set persistence is then used to provide a computational summary of the topological structure of the KDE.  We give a brief explanation of persistent homology in Section~\ref{sub-level} and the procedure for applying TDA to the $p$-$q$ data in Section~\ref{1D}. The results are explained in Section \ref{results} while case studies on the Rossler system and the Logistic map are given in Sections~ \ref{rossler_system} and~\ref{Logistics} respectively. The implementation of this method is  described with Algorithm~\ref{alg:TDA_test} which we implement in MATLAB\textsuperscript{\textregistered} and is published online~\cite{Tempelman2019}.
}
\subsection{Sub-level Set Persistence}
\label{sub-level}

Only a brief explanation of persistent homology, commonly known as persistence, will be provided in this paper. For a more comprehensive explanation, the reader is directed to~\cite{Munkres1993, Munch2017, Cohen-Steiner2006, Ghrist2014, Edelsbrunner2013, Edelsbrunner2008}. 

\begin{figure}[h!]
	\centering
	\includegraphics[width = \textwidth]{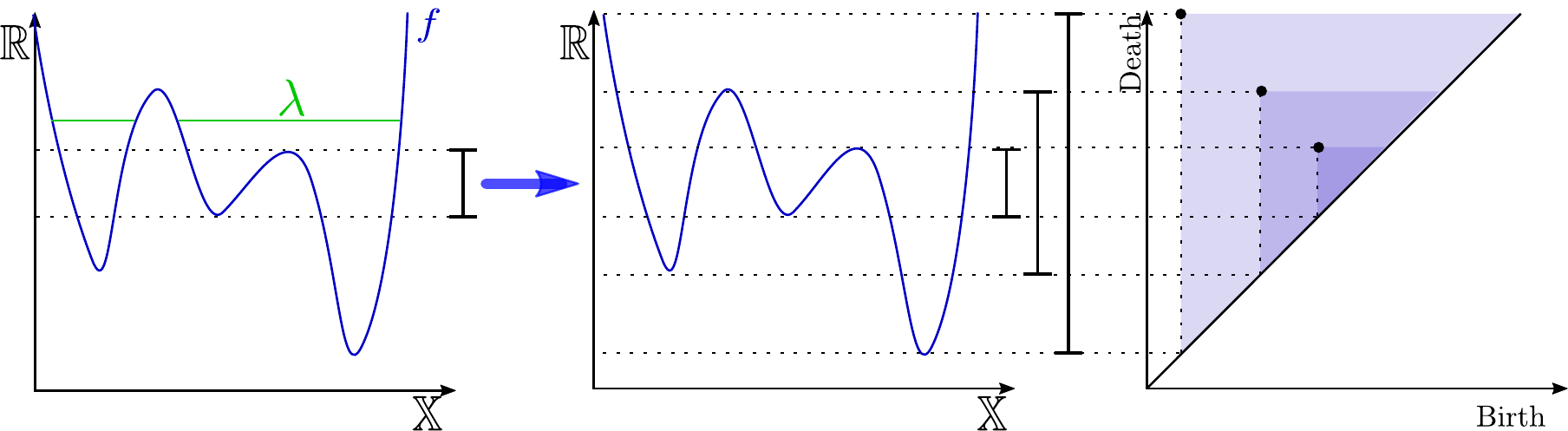}
	\caption{An example showing the construction of the 0D persistence diagram of the function $f:{\mathbb{X}}\to{\mathbb R}$. (left) The rising ``water'' level on cross-section cut of a function $f$ in topological space ${\mathbb{X}}$ and (middle) the sub-level sets produced by tracking the ``birth''  and ``death'' of topological features on $f$. The birth and death times are plotted against on each other (right) to produce the corresponding persistence diagram for the function. }
	\label{fig:sublevel}
\end{figure}

\begin{figure}[h!]
	\centering
	\includegraphics[width=0.75\linewidth]{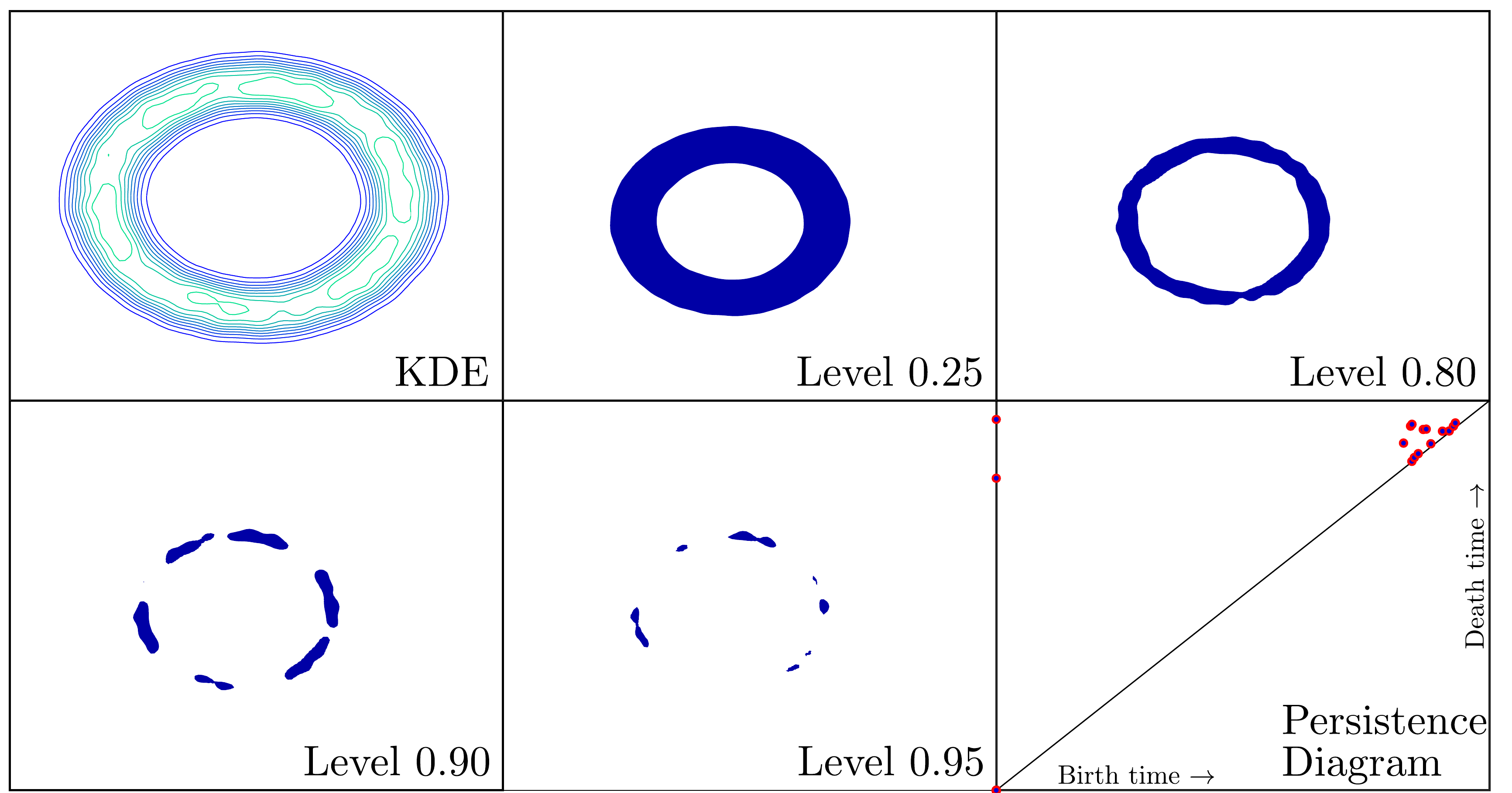}
	\caption{Sub-level filtrations for the KDE (left) of a $p$-$q$ projection for a periodic time series and the resulting persistence diagram for the 0D persistence.}
	\label{fig:sublevel_1D}
\end{figure}

Persistence can be applied to 3-dimensional data such as contour plots or even gray-scaled images. Each point in topological space $\mathbb{X}$ is assigned a real number, which for the purpose of this explanation, may be thought of as a height coordinate.
This resulting function $f:\mathbb{X}\to\mathbb{R}$ can have local maxima and minima. 
Persistent homology is interested in changes in topology at various {\it sub-level sets}.
For this function $f$, the  $\lambda$ sub-level set is
\begin{equation}
L_{\lambda} = \left\{x:f\left(x\right)\leq\lambda\right\}=f^{-1}\left(\left[-\infty, \lambda\right]\right).
\label{eq:sublevel}
\end{equation}
For any pair of levels $\lambda_1>\lambda_2$ the condition  $L_{\lambda_1}\supseteq L_{\lambda_2}$ must be satisfied and the collection of sets $\{L_{\lambda}\}_{\lambda \in{\mathbb R}}$ forms a filtration where the level is the index set~\cite{Berry2018}. 
Consider Fig.~\ref{fig:sublevel} which shows a toy example for computing 0D sub-level persistence for a function $f$. One way to visualize the process is to imagine a rising water level  beginning at the base of $\mathbb{X}$ and account for each time a new minimum of the surface is touched and new maximum is reached (see Fig.~\ref{fig:sublevel}).
As the level $\lambda$ is raised, new topological features are captured through generators of its homology group. 
Two dimensions of homology groups are considered. The 0D groups capture connected components (see Fig.~\ref{fig:sublevel}) while the 1D groups captures regions forming circular structures (see Fig.~\ref{fig:sublevel_1D}).

As new levels are introduced, new generators of homology groups are formed and existing generators merge. The level at which these generators and groupings occur are the {\it birth} and {\it death} time, respectively, of the topological feature. Therefore, the three characteristics for each generator are the homology dimension, birth time, and death time.
The collection of these points is drawn as what is referred to as a persistence diagram. For $|D|$ generators, the persistence diagram is ${\mathsf{D}} = \left\{(r_j,b_j,d_j):j=1,\dots,|D|\right\}$ where $r_j$, $b_j$, and $d_j$ are, respectively, homology order, birth time, and death time of the $j$th generator~\cite{Berry2018}.   

\subsection{1D Persistence of Kernel Density Estimates}
\label{1D}
\begin{figure}[h!]
	\centering
	\includegraphics[width = \textwidth]{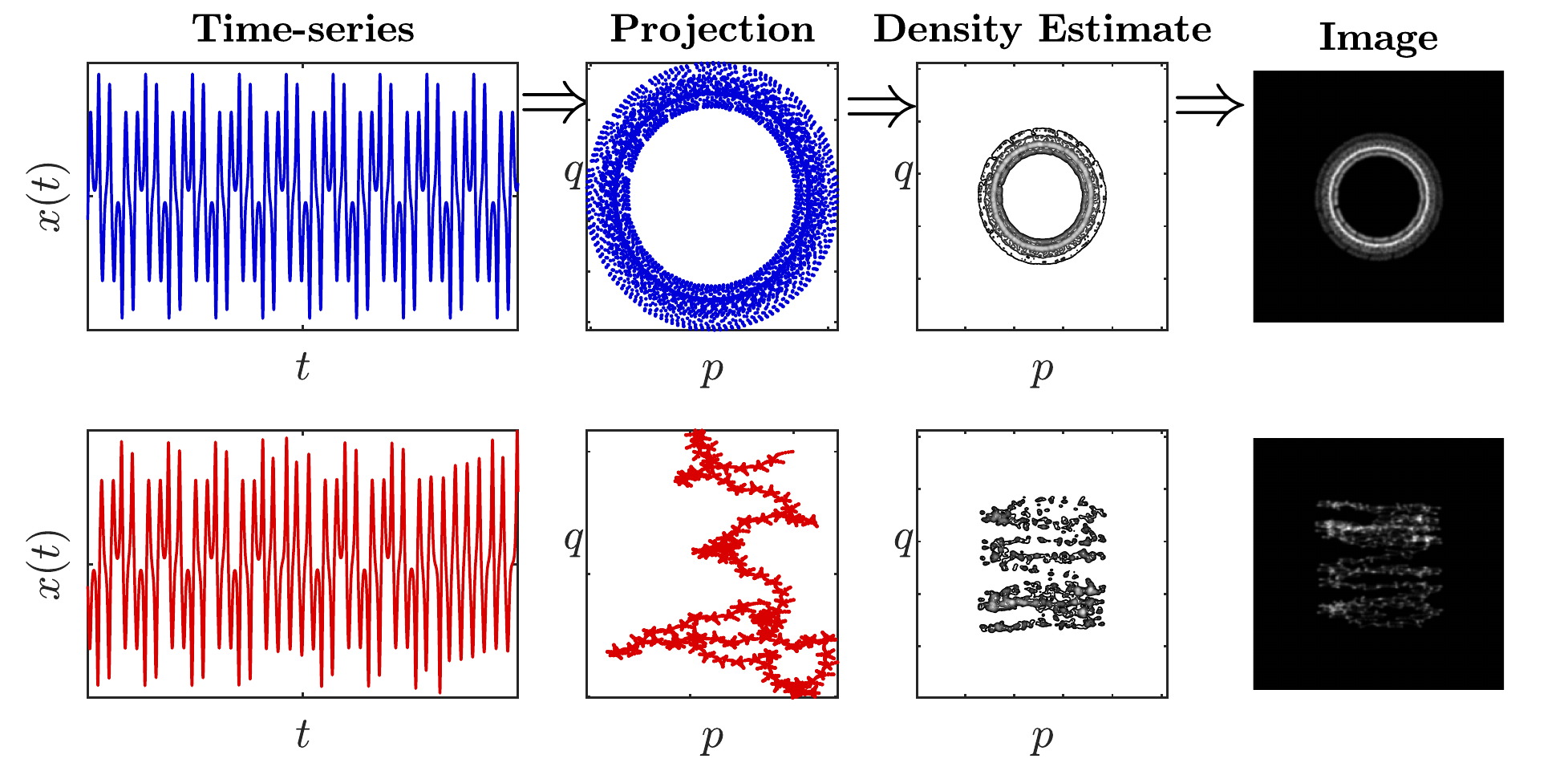}
	\caption{The transformation from time series data to a gray-scaled image. The raw-time series (left) are projected into the $p$-$q$ plane (middle-left) before a bi-variate kernel density estimate produces a smooth topological surface (middle-right). The density estimates are then turned into gray-scale images (right) for sub-level set persistence. Shown here is one iteration for laminar flow (top) and chaos (bottom).}
	\label{fig:method}
\end{figure}
%

\textcolor{black}{
	In this paper, we utilize the 0D  sub-level set persistence to extract the underlying characteristics of the probability densities of the $p$-$q$ projections. 
	The  KDEs correspond to the $p$-$q$ data according to~\cite{Botev2010}
	\begin{equation}
	\hat{f}({x}, {\bm { \mathrm H}}) = \frac{1}{n}\sum_{i=1}^{n}{\textbf{K}_{\mathrm H}}\left({x}-{x}_i\right),
	\label{eq:BVKD}
	\end{equation}
	where ${x} = (p,q)$, $n$ is the number of elements in $\{x_1,x_2,\dots,x_n\}$, and ${\textbf{K}_{\mathrm H}}$ is the kernel function. 
	The resulting KDEs are converted to a gray scale image. The gray scale image is normalized to have a pixel value in [0, 1].   
}

\textcolor{black}{
	It should be noted that increasing the grid resolution will drastically lengthen the runtime of the program.   To reduce computation time of the corresponding sublevel set persistence, a KDE grid size of $2^5\times2^5$ is used. 
	This is a relatively coarse grid resolution and results in KDEs which are jagged when compared to a KDE of a higher resolution. This could lead to misleading computational summaries of the KDE topology, thus a 2D Gaussian smoothing function 
	\begin{equation}
	G_{h} = \frac{1}{2\pi h^2}e^{-\frac{x^2+y^2}{2h^2}}
	\label{eq:Gaussian_filter}
	\end{equation}
	is used. We found $h=1.3$ to be a good choice for the smoothing kernel width at a $2^5\times2^5$ grid resolution (the effect of this kernel width on the results of the persistence test is discussed in Section~\ref{Kernel}).
	The DIPHA\footnote{https://github.com/DIPHA/dipha}  library was used to generate all the needed 0D persistence diagrams.
}
\subsection{Results and Discussion}
\label{results}

\begin{figure}[h!]
	\begin{subfigure}{\textwidth}
		\centering
		\includegraphics[width = .65\textwidth]{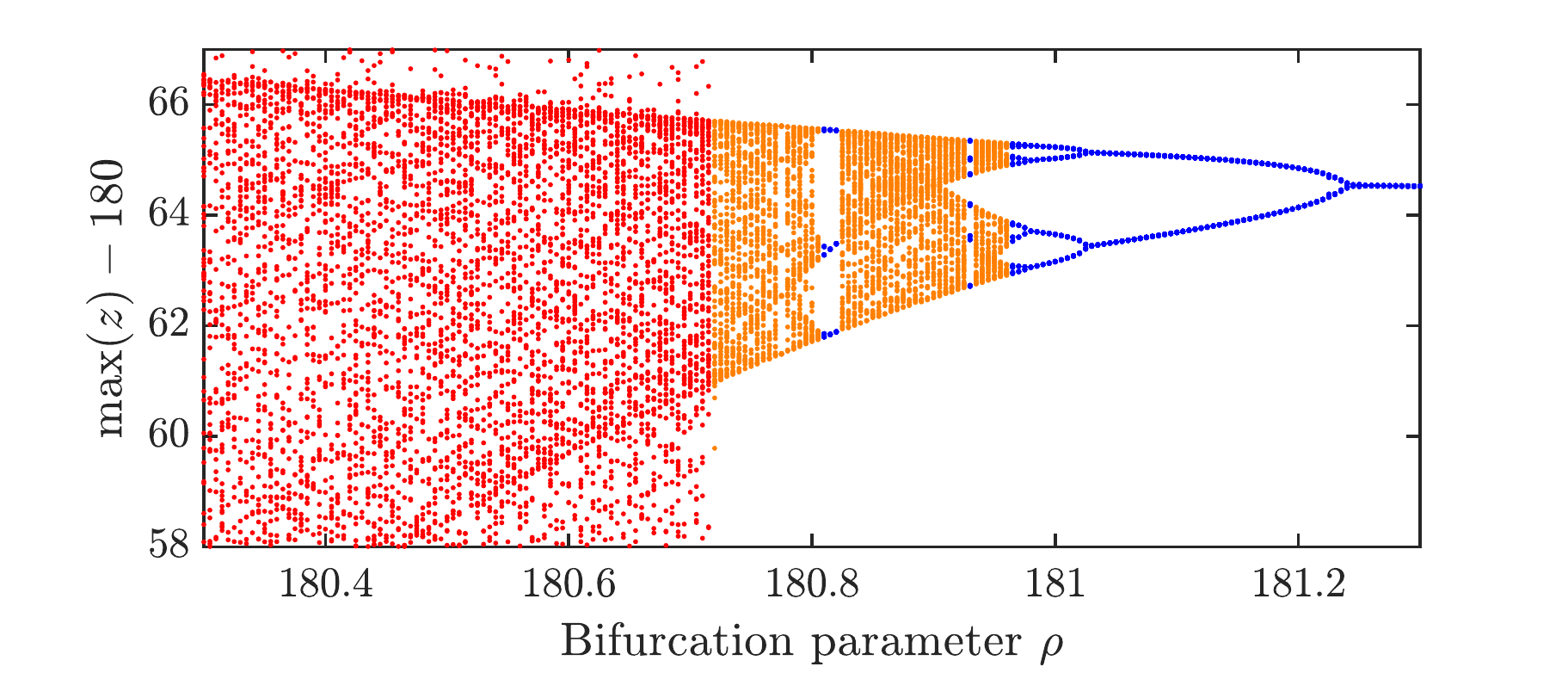}
	\end{subfigure}
	\begin{subfigure}{\textwidth}
		\centering
		\includegraphics[width = \textwidth]{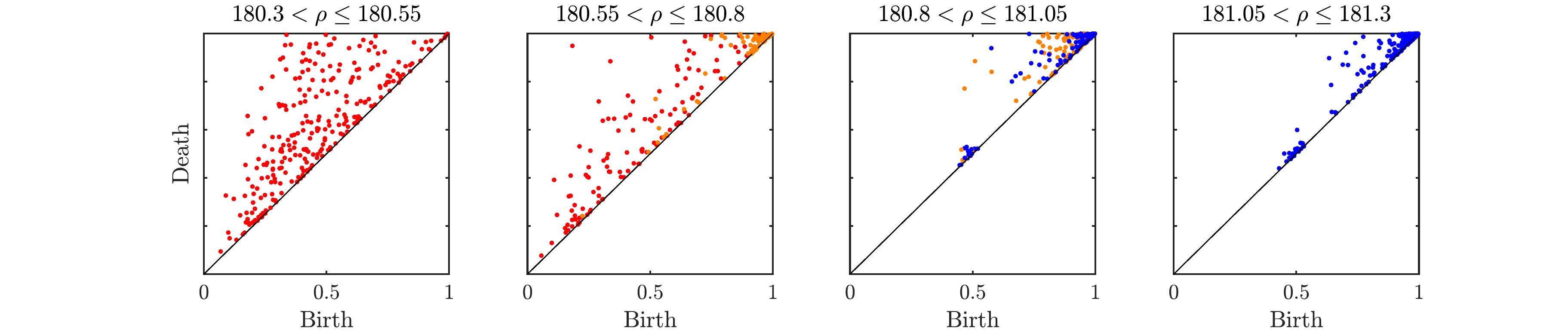}%
	\end{subfigure}
	\caption{(top) The bifurcation diagram of the Lorenz model as the bifurcation parameter $\rho$ is varied between 180.3 and 181.3. (bottom) The persistence diagrams generated from kernel density estimates of $p$-$q$ projections acquired from time series at ranges of the bifurcation parameter. The color scheme corresponds to the underlying dynamics as found in~\cite{Wernecke2017} where blue indicates periodicity, orange indicates PPC, and red indicates chaos.  }
	\label{fig:birth_death}
\end{figure}

The central motivation for this study is to yield a reliable interpretation of the underlying dynamics of time series data using persistence. To better understand how this may be achieved, refer to Fig.~\ref{fig:birth_death}. 
Here, multiple persistence diagrams are overlaid for ranges of the bifurcation parameter $\rho$ of the Lorenz model. 
The persistence points with a birth time near zero are omitted from Fig.~\ref{fig:birth_death} since they are considered noise in this application. The persistence points shown in Fig.~\ref{fig:birth_death} show that when persistence diagrams are generated from time series associated with a periodic bifurcation parameter, the persistence points accumulate in the upper-right corner of the persistence diagram.

\textcolor{black}{
	These produce birth/death combinations which are tightly clustered and bounded to regions on or near the diagonal. There is an apparent correlation here between the birth and death times of the persistence points. 
	Furthermore, these points all appear to be occurring at higher birth and death times than what is seen for their chaotic counterparts. 
	The persistence data generated form PPC trials are visually indistinguishable from periodic trials. 
	On the other hand, chaos based persistence diagrams are visually perceptible from periodic when many trials are overlaid as shown in Fig.~\ref{fig:birth_death}. 
	Thus, the hypothesis to test is whether or not this behavior may be reliably used to trace back the dynamics of the original time series.  
}

{\bf Scoring}:  \textcolor{black}{Infinitely many $p$-$q$ projections may be generated for a given time series since $c$ may be indefinitely drawn from $(0,\pi)$. }
Thus, we use an ensemble of $p$-$q$ projections and use a statistical summary of birth and death times to deduce the original dynamics of the time series. 
In this section, the mean distance of persistence points from the diagram's origin are used to discern the underlying dynamics of the time series.

\begin{figure}[t!]
	\centering
	\begin{subfigure}{\textwidth}
		\centering
		\includegraphics[width = \textwidth]{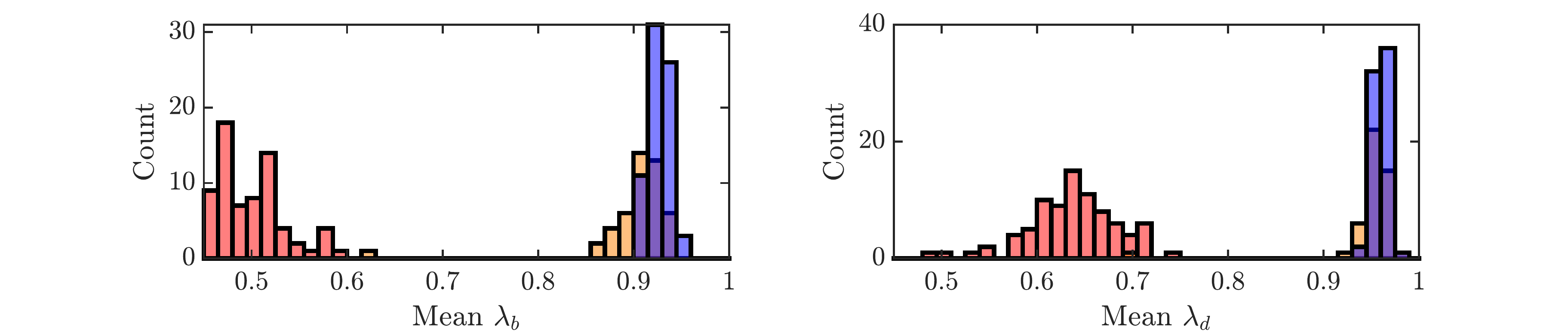}
	\end{subfigure}
	\caption{Basic statistical summary of 200 persistence diagrams taken from the gray-scaled image renditions of the KDE of the $p$-$q$ projections for 200 simulations of the Lorenz system. The red bars correspond to chaos, the orange bars correspond to PPC, and the blue bars correspond to periodic dynamics.  200 simulations are used for $\rho\in[180.3, \ 181.3]$ to generate the full range of dynamics. The images were smoothed by Gaussian filtering and the data used for the statistical summary ignores points with birth very close to the origin.  }
	\label{fig:stats}
\end{figure}
\begin{figure}[h]
	\centering
	\includegraphics[width=.9\textwidth]{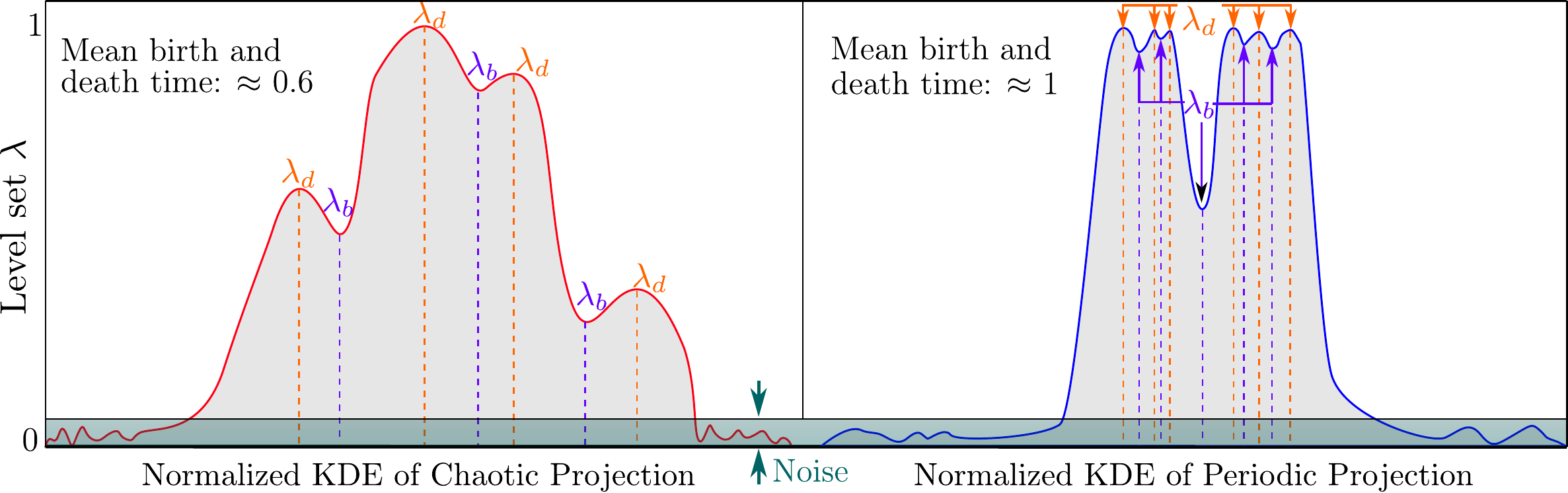}
	\caption{Visual interpretation of sub-level filtrations for the normalized KDE  of a $p$-$q$ projection for a (left) chaotic and (right) periodic time series and the resulting 0D persistence.}
	\label{fig:KDE_dgm}
\end{figure}
\textcolor{black}{
	The threshold for chaotic versus periodic or PPC behavior in persistence diagrams can be defined as a result of the location of the points in the diagram.
	A basic statistical evaluation of the trials used in Fig.~\ref{fig:birth_death} is used to elucidate how such a threshold is defined. 
	Looking at the average birth and death times, Fig.~\ref{fig:stats} shows that the average time of birth and death for periodic data is substantially higher than the chaotic counterpart (approximately $0.90$--$1.0$ for periodic and $0.40$--$0.65$  for chaotic).
	The statistical summary outlined in Fig.~\ref{fig:stats} ignores the birth/death combination on or near zero birth time vertical. 
	These persistence points do not reveal any relevant information, and can be thought of as irrelevant noise in the KDE plots. 
}
\textcolor{black}{
	The statistical summary was conducted by taking the mean birth and death time of each diagram for 200 realizations of $p$-$q$ projections at each bifurcation parameter and then computing the ensemble mean of these values. 
	The average birth and death times are substantially higher for periodic and PPC data; therefore, we score the data using the average distance to the origin according to
	\begin{equation}
	PS_1 =\Bigg \langle \sum_{j=1}^{N_i} \left\{\frac{\ \sqrt{(d_j^2 + b_j^2)}}{N_i}, (b_j,d_j)\in {\mathsf D}_i\right\} \Bigg \rangle, 
	\label{eq:PS1}
	\end{equation}
	where $PS_1 \in [0,\sqrt{2}]$, ${\mathsf D}_i$ is the persistence diagram generated form the $i$th selection of $c\in(0,\pi)$, and $N_i$ is the number of 0D persistence points in ${\mathsf D}_i$.
}

\textcolor{black}{
	The reason for the higher average birth and death times in the persistence diagrams for KDEs generated from periodic time series as compared to chaotic ones is illustrated in Fig.~\ref{fig:KDE_dgm}. The diffusive nature of the $p$-$q$ trajectories~\cite{Gottwald2016} for chaotic data results in KDEs which typically have multiple local maxima and minima randomally appearing in the $p$-$q$ space. These local extrema correlate to the births and deaths of homology classes in the persistence diagram. 
	Alternatively, the bounded projections of periodic time series data will consistently produce a KDE which is bound to a finite region in the $p$-$q$ space with prominent peaks representing frequently visited parts of the $p$-$q$ space. This structure of the KDE leads to the homology classes being constant during the filtration until the top of the structure is reached. 
	At which point, ridges near the peak of the KDE produce many generators near a value of one which die soon after birth. 
}

\begin{figure}[t]
	\centering
	\includegraphics[width=\textwidth]{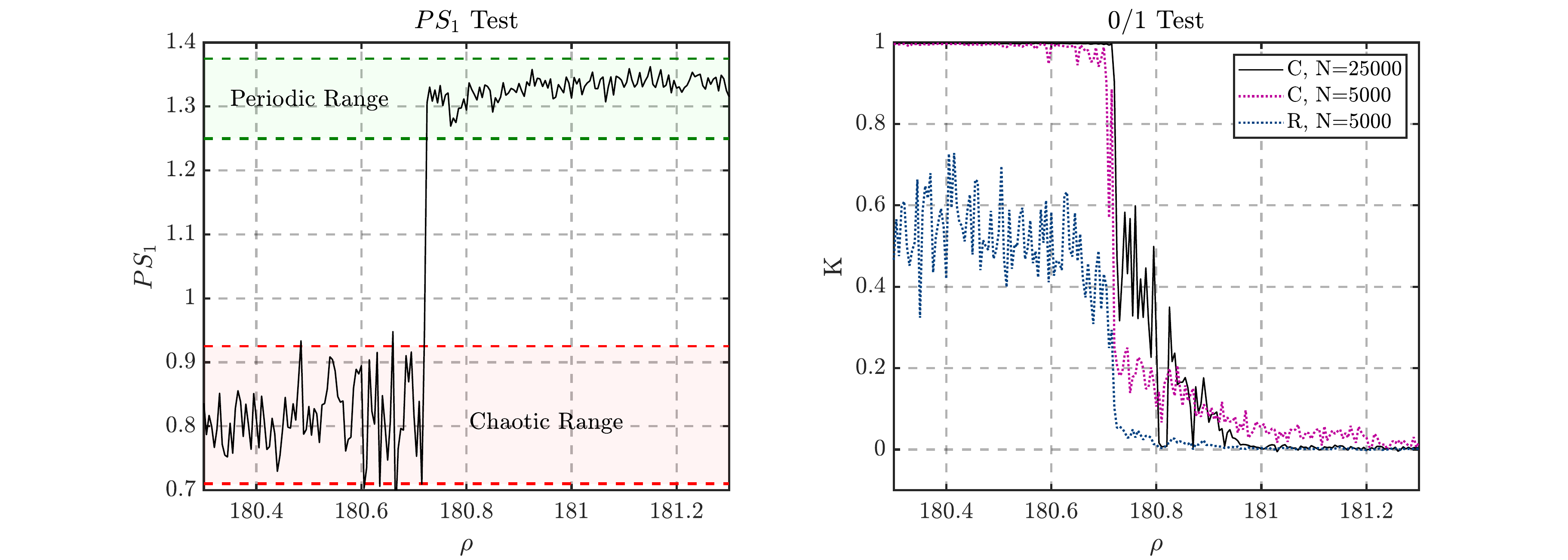}
	\caption{A direct comparison (left) the proposed method of averaging persistence data based on its position with respect to the origin and (right) the Gottwald 0--1 correlation (C) and regression (R) tests. The long--time 0--1 correlation score (N=25000) is provided as a reference.}
	\label{fig:chaostest_lorenz}
\end{figure}
Thus, it can be concluded that for the Lorenz system, if the $PS_1$ score for a set of $p$-$q$ projections is near $\sqrt{2}$, then the underlying dynamics are periodic. 
The mean distance from the origin $PS_1$ was consistently found to be a successful score for classifying chaos for the systems studied in this paper and thus it will be the focus of the results section. 

\begin{algorithm}[t]
	\caption{Testing for chaos with persistence}\label{alg:TDA_test}
	\begin{algorithmic}[1]
		\Procedure{$PS_1$}{$\Phi$,$f_s$}								\Comment{{\it time series data matrix}}
		\State $\phi\gets $subsample($\phi,f_s$) 						\Comment{{\it Select observable and sub-sample}}
		\For{$i =$ $1:N$}												\Comment{{\it $N=200$ is suggested}}
		\State $c_i \gets $rand$(0, \pi) $								\Comment{{\it Random c value}}
		\State $[{\bf p},{\bf q}] \gets f({\bf x},c)$					\Comment{{\it Project into $p$-$q$ space}}
		\State ${\bf P} \gets f({\bf p},{\bf q})$						\Comment{{\it Get bi-variate density estimate of (p,q)}}
		\State ${\bf I} \gets {\bf P}$									\Comment{{\it Convert to image}}
		\State $[{\bf B}, {\bf D}, {\bf d}] \gets$ DIPHA(${\bf I}$)		\Comment{{\it Call DIPHA and get persistence data}}
		\State ${\bf B}\gets {\bf B}({\bf d}=0)$						\Comment{{\it Cut out persistence data with dimension 1}}
		\State ${\bf D}\gets {\bf D}({\bf d}=0)$
		\State ${\bf s}\gets$ $f({\bf B},{\bf D})$						\Comment{{\it Calculate distance from origin for each point}}
		\State ${\bf S}_i\gets$ mean(${\bf s}$)							\Comment{{\it Take the mean of distances for the diagram}}
		\EndFor
		\State \textbf{end}
		\State \textbf{return} mean($\bf S$)							\Comment{{\it Compute ensemble mean on index $i$}}	
		\EndProcedure
	\end{algorithmic}
\end{algorithm}

\begin{figure}[t!]
	\centering
	\begin{subfigure}{\textwidth}
		\centering
		\includegraphics[width=\textwidth]{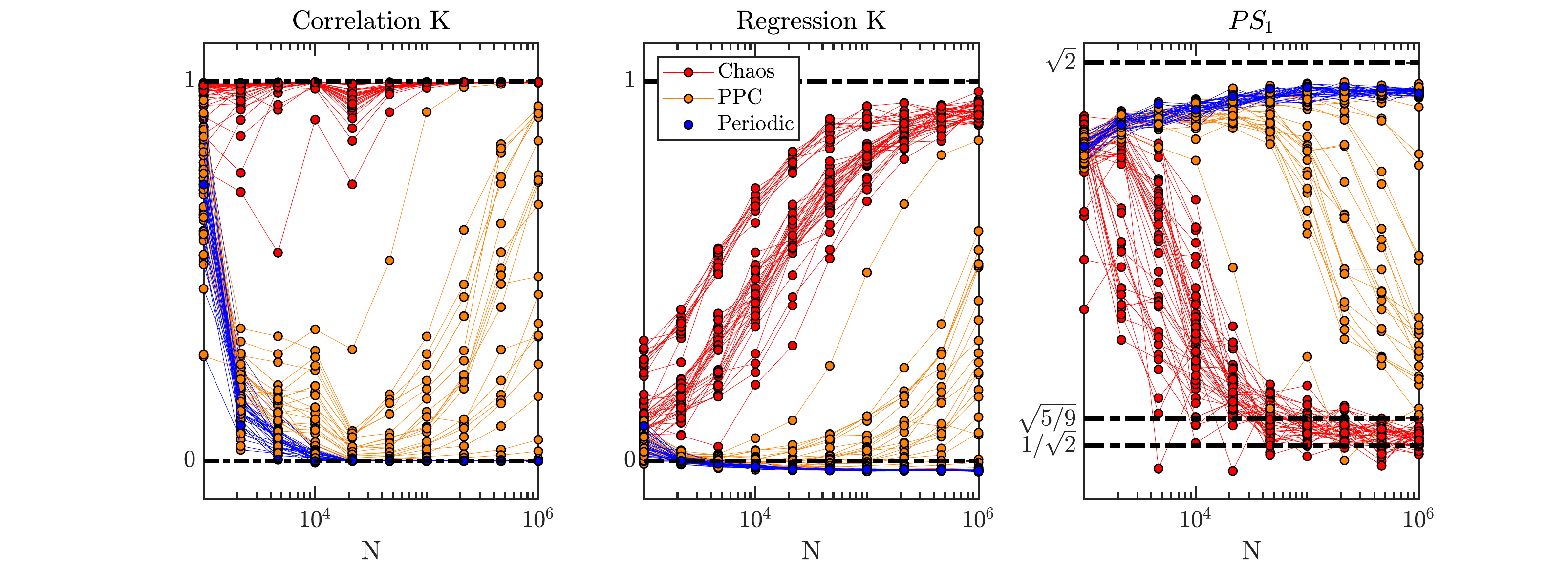}
	\end{subfigure}%
	\caption{The convergence of (left) the 0-1 correlation test, (middle) the 0--1 modified regression test and (right) the $PS_1$ scores. }
	\label{fig:conv}
\end{figure}

\textcolor{black}{
	Figure~\ref{fig:chaostest_lorenz} shows a direct comparison between our proposed method (see Fig.~\ref{fig:method} and Algorithm~\ref{alg:TDA_test}) and the 0--1 correlation test. 
	The 0--1 test and $PS_1$ scores are performed using $N=5000$ (after sub-sampling), and a long time limit ($N=25000$) evaluation of the 0--1 test is included as a reference case~\cite{Gottwald2016,Gottwald2008}. 
	When the dynamics are periodic ($\rho>180.73$), the distance from the origin stabilizes near 1.3. 
	When chaotic dynamics are present, the $PS_1$ score stabilizes at approximately 0.8. 
	The scores of 1.3 for periodic and 0.8 for chaotic were found to be consistent when testing with longer time series as well.
}

\textcolor{black}{
	The numerous numerical experiments that we preformed seemed to yield $PS_1$ scores close to the $[1/\sqrt{2}\approx 0.7$, $\sqrt{5/9}\approx 0.75]$. 
	These two scores coincide with the geometric centers of the diagonal line and the upper left triangle of the persistence diagram, respectively. 
	This suggests that the persistence of the corresponding probability density estimates of the projections possess a global shape which appears similar to the one obtained form the 0D persistence of Gaussian random field excursions \cite{Adler2019}. 
	However, computing distributions of persistence diagrams is not a trivial task~\cite{Adler2010, Adler2014, Kahle2013,Adler2019} and is complicated by factors such as the highly nonlinear nature of the resulting metric space, as well as the non-uniqueness of geodesics and means in the persistence diagram space. 
	Finding a closed-form expression for these expectations is a topic for future research. 
}

\textcolor{black}{
	Figure~\ref{fig:chaostest_lorenz} also shows how the 0--1 test and the $PS_1$ scores behave when PPC dynamics are present in the bifurcation parameter range $\rho\in[180.7, 181]$. Specifically, in this parameter range both methods return a periodic diagnostic for $N = 5000$. 
	However, Fig.~\ref{fig:conv} shows that the long-time limit ($N =10^6$ points) for PPC data deviates from this periodic result for both the 0--1 correlational and regression scores.
	The figure shows that as the time series is extended, both the 0--1 correlation test and the $PS_1$ statistic start to yield non-binary results between zero and one, but not necessarily near zero or one. Although this is found to be consistent in both the modified regression and correlation 0--1 tests, the divergence from a periodic score occurs quicker with the latter.
	Similarly, a $PS_1$ score between 0.8 and 1.3, but not necessarily near 0.8 or 1.3, is returned from the $PS_1$ statistic. 
	\textcolor{black}{ One may argue that this is simply an issue of finite time. 
		While it appears that the PPC data are tending towards one of the two diagnostic limits for both the 0--1 tests and $PS_1$ scores, it does so at too slow of a rate for convergence to be achieved without resorting to excessively taxing computations. }
	In contrast, scores for strongly chaotic or fully periodic regimes remain consistent during this long-time evaluation. 
	This shows that it is difficult to distinguish PPC solely from $p$-$q$ projections. 
}

\begin{figure}[b!]
	\centering
	\begin{subfigure}{\textwidth}
		\includegraphics[width=\textwidth]{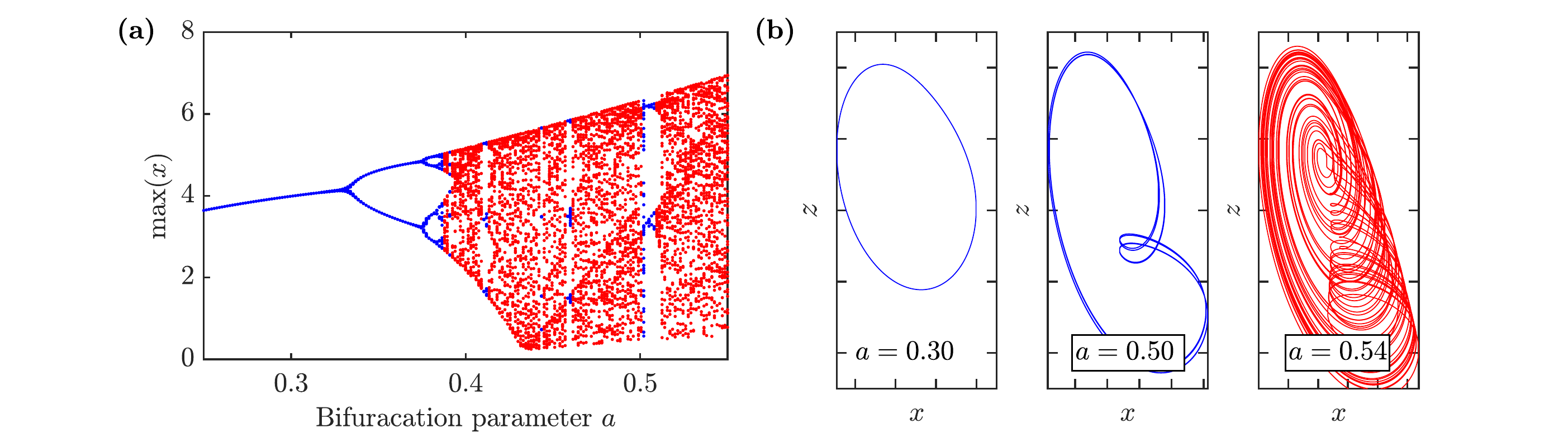}
	\end{subfigure}
	\begin{subfigure}{\textwidth}
		\includegraphics[width=\textwidth]{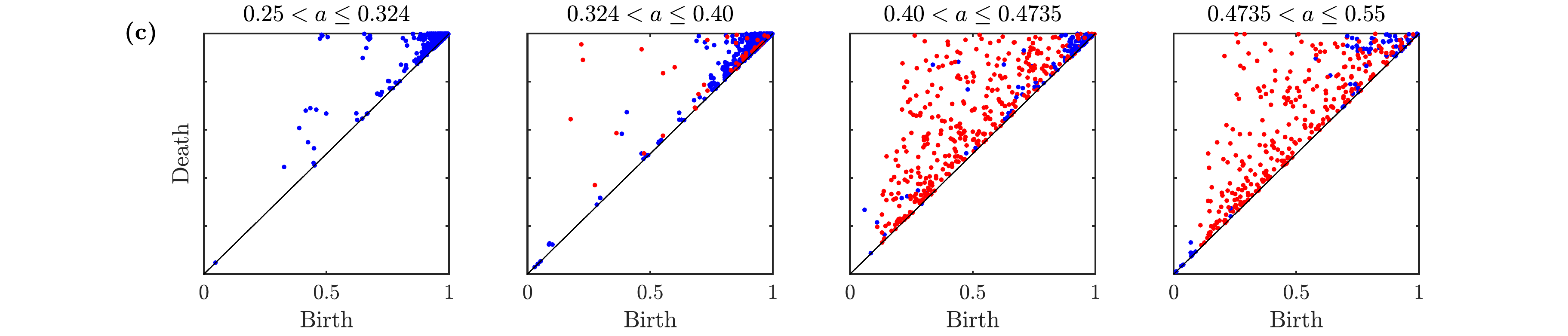}
	\end{subfigure}
	\begin{subfigure}{\textwidth}
		\includegraphics[width=\textwidth]{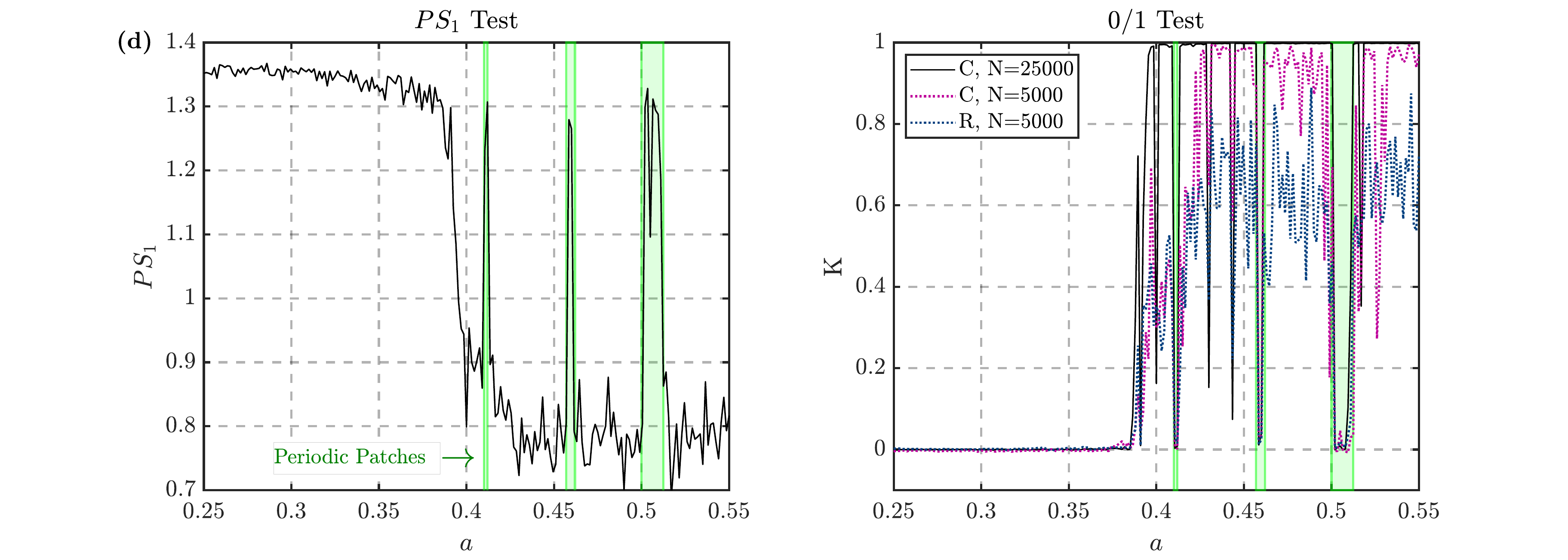}
	\end{subfigure}
	\caption{(a) The bifurcation diagram of the Rossler model for $a\in[0.25, 0.55]$, (b) 2-D projection of the phase space in the $x$-$y$ plane for multiple values of the bifurcation parameter, (c) resulting persistence diagrams generated over ranges of $a$, and (d) the results $PS_1$ and 0--1 correlation (C) and regression (R) testing. The long-time correlation (N=25000) 0--1 score is provided as a reference. }
	\label{fig:rosslerbif}
\end{figure}

\subsubsection{Case Study: The Rossler System}
\label{rossler_system}
\textcolor{black}{
	In order for our method to be applicable to a wide class of signals, it must yield consistent numerical results for periodic and chaotic time series regardless of the source of the signal. To investigate this, we use the Rossler system to gather additional synthetic data.
	Like the Lorenz equations, the Rossler model is a classic system which has been studied widely~\cite{Roessler1976} and it is described by
	\begin{equation}\label{eq:rossler}
	\dot{x} = -y-z, \hspace{15pt} \dot{y} = x+ay, \hspace{15pt} \dot{z} = \gamma + z(x-c),
	\end{equation}
	where $a$, $c$ and $\gamma$ are scalars. We only consider fully periodic and fully chaotic signals in this section. 
	The Rossler data is generated through numerical simulation using the MATLAB\textsuperscript{\textregistered} ODE45 integration tool with $\Delta{t} = 0.01$, a relative tolerance of $10^{-5}$, and a transient cut-off of 1,000 seconds. 
	The parameters selected for this study are $a\in[0.25\hspace{4pt} 0.55]$, $\gamma=2$, and $c=4$~\cite{Letellier1995}. 
	Figure~\ref{fig:rosslerbif} shows the results of the 0--1 correlation test along with the $PS_1$ scores for $N=5000$ as well as the long time reference using $N=25000$ using the 0--1 correlation test. The bifurcation diagram and trajectories in the phase space are also given in Fig.~\ref{fig:rosslerbif}a and \ref{fig:rosslerbif}b, respectively.}

\begin{figure}[b!]
	\centering
	\begin{subfigure}{\textwidth}
		\includegraphics[width=\textwidth]{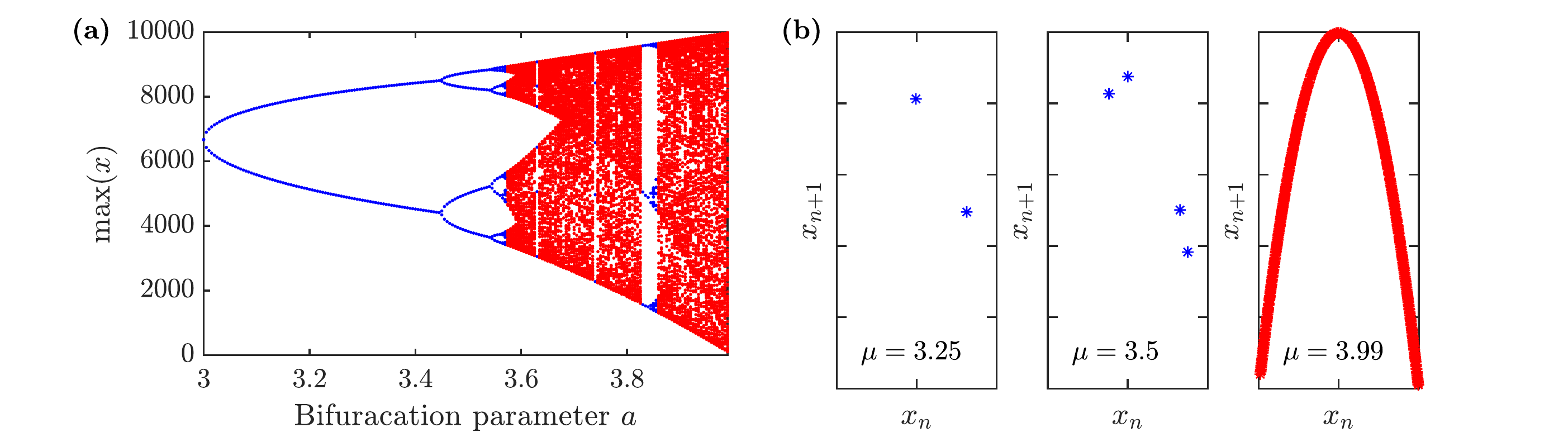}
	\end{subfigure}
	\begin{subfigure}{\textwidth}
		\includegraphics[width=\textwidth]{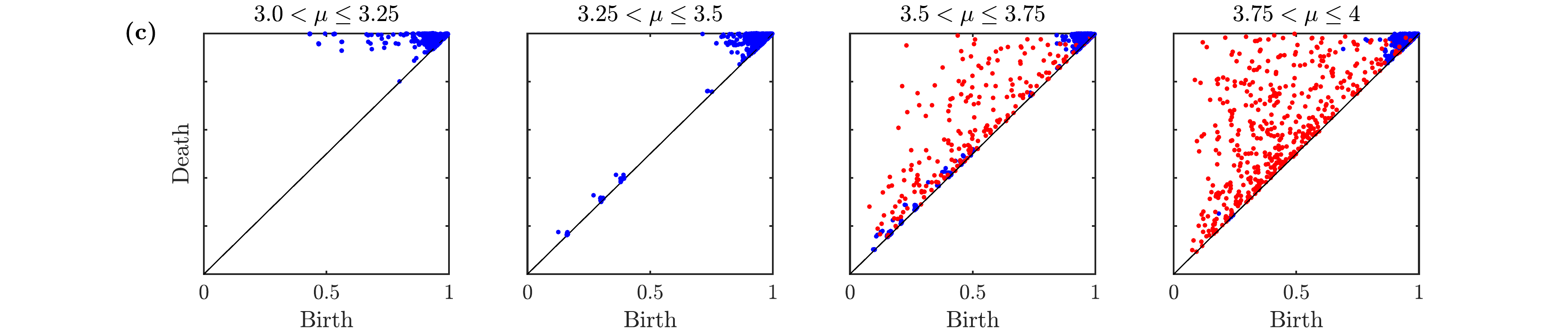}
	\end{subfigure}
	\begin{subfigure}{\textwidth}
		\includegraphics[width=\textwidth]{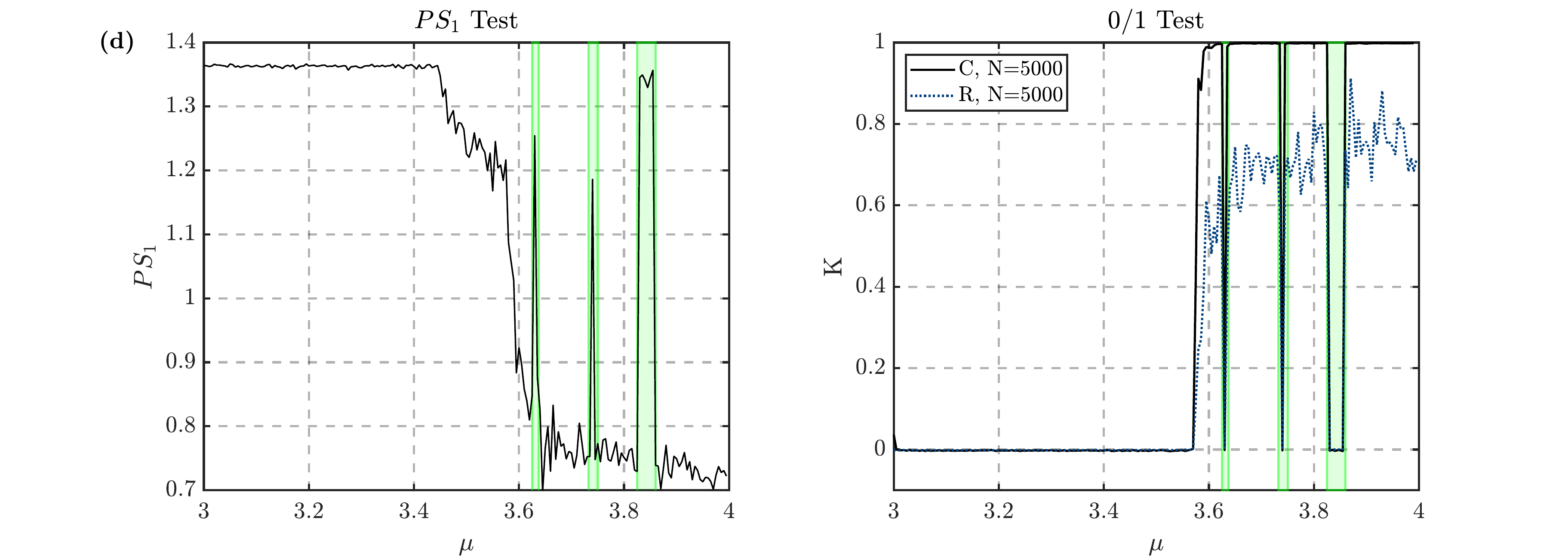}
	\end{subfigure}
	\caption{(a) The bifurcation diagram of the Logistics map for $\mu\in[3,4)$ (b) the plots of $x_n$ versus $x_{n+1}$ as $\mu$ is varied, (c) persistence diagrams generated at different ranges of the bifurcation parameter, and (d) the resulting $PS_1$ and 0--1 correlation (C) and regression (R) scores. The long-time correlation (N=25000) 0--1 score is provided as a reference. }
	\label{fig:logbif}
\end{figure}

The same TDA based approach we used for the Lorenz system for chaos detection is applied to the Rossler data.
Time series were generated for a span of bifurcation parameters. \textcolor{black}{The minimum significant frequency sub-sampling method described in section~\ref{a_tda_approach} (i.e. $2f_{\rm max}<f_s<4f_{\rm max}$) was applied to the Rossler simulations which are then projected into the $p$-$q$ space per Eq.~\eqref{eq:pq}.}
The 0D sub-level set persistence then generated the birth/death times for gray-scale renditions of the normalized KDE of these projections which were smoothed per Eq.~\eqref{eq:Gaussian_filter} with $h=1.3$. 
\textcolor{black}{
	The $PS_1$ and the correlation 0--1 scores were obtained for the time series generated over the different values of $a$ with $N=5000$.
	Figure~\ref{fig:rosslerbif} shows that the correlation 0--1 test and the $PS_1$ scores perform comparably for $N=5000$, with a more definitive classification  $PS_1$ scores for $a\in[0.375,0.425]$. In this range of $a$ values, the regression 0--1 test values are rather ambiguous since the $K$ values commonly fall between 0.5 and 0.8 over these selections of $a$.   
}

\textcolor{black}{
	On the other hand, the $PS_1$ and 0--1 scores do not consistently yield a definitive periodic or chaotic diagnostic for $a\in[0.4, 0.55]$. While there is a ``gray area" in this parameter range, Fig.~\ref{fig:rosslerbif} shows the emergence of elevated plateaus in the $PS_1$ scores and 0--1 correlation scores for $a>0.38$. These values coincide with windows of periodicity shown in the bifurcation diagram and in the long--time reference scores of the 0--1 correlation test.  
	Such gray areas are to be expected when dealing with finite time series since the length of the time series may not be sufficient to capture the full dynamics of the attractor~\cite{Gottwald2016,Gottwald2008}. 
}

\subsubsection{Case Study: The Logistic Map}
\label{Logistics}

\textcolor{black}{
	We further test our method on the Logistic map which is an iterative map governed by~\cite{May2004}
	\begin{equation}
	x_{n+1} = \mu x_n(1-x_n).
	\label{eq:logistics}
	\end{equation}
	This nonlinear map was introduced as  a simplified population model where $x_n$ is a ratio between existing population and maximum population. Here, the bifurcation parameter is $\mu$ with transitions in the dynamics occurring during $\mu\in [3,4)$. 
}

\textcolor{black}{
	Both the 0--1 correlation test and the persistence scoring method are applied to time series generated from Eq.~\eqref{eq:logistics} over the range of $\mu$ values. Like before, the kernel width $h$ in Eq.~\eqref{eq:Gaussian_filter} is set to 1.3 and the persistence points found very close to the origin are ignored. 
	The 0--1 correlation test does a very good job of correctly separating regions of chaos and periodicity through nearly perfect binary scores. 
	The $PS_1$ scores do not separate as cleanly as shown in Fig.~\ref{fig:logbif}.
	However, the transition from periodic behavior to chaotic behavior ($\mu \approx 3.55$) as well as the pockets of periodicity in $\mu\in[3.55,4)$ are clearly noticeable in the persistence scores.  
	Additionally, Fig.~\ref{fig:logbif} shows that $PS_1\approx1.4$ indicates a periodic time series while $PS_1\approx0.80$ indicates chaos in the Logistic map.
	This is consistent with the results in Fig.~\ref{fig:chaostest_lorenz} for the Lorenz system and in Fig.~\ref{fig:rosslerbif} for the Rossler system.   
}

\begin{figure}[h!]
	\centering
	\begin{subfigure}{\textwidth}
		\centering
		\includegraphics[width=.65\textwidth]{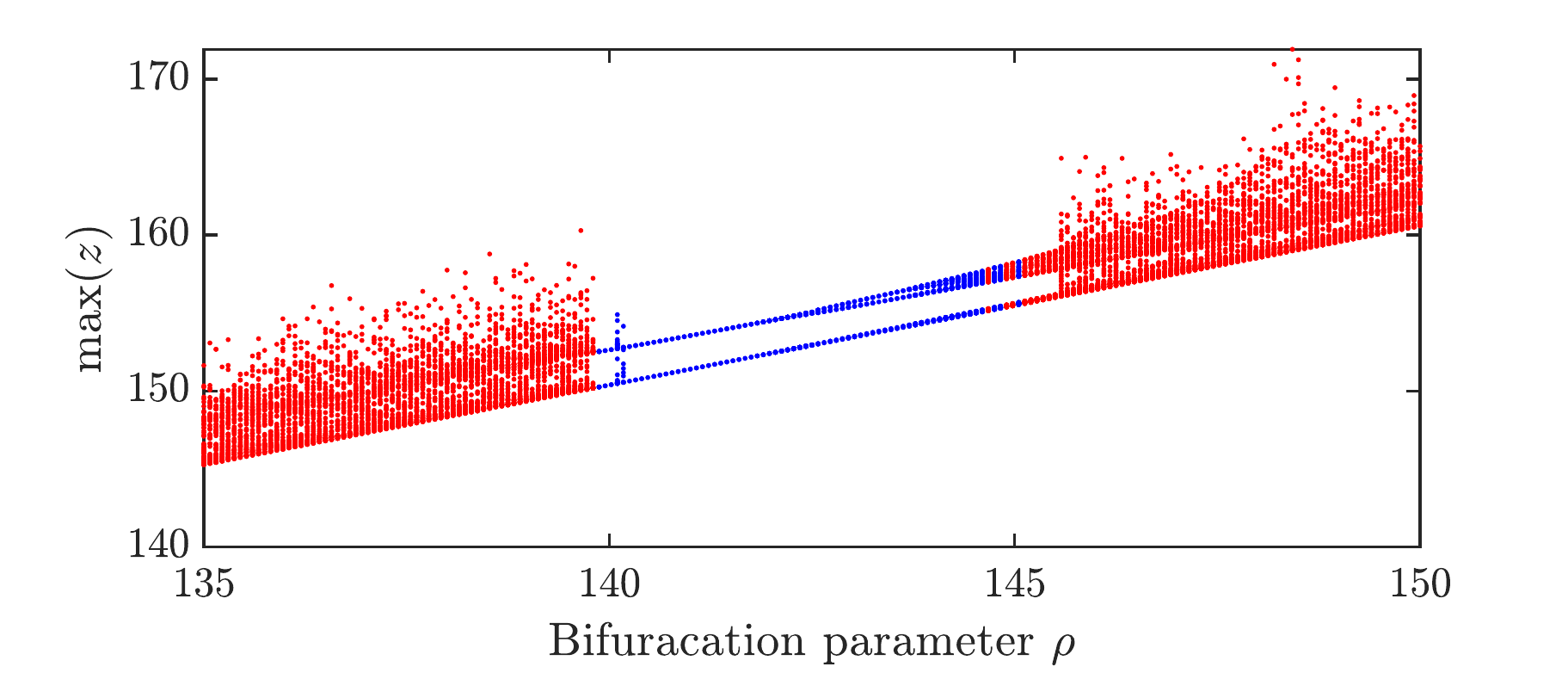}
	\end{subfigure}
	\begin{subfigure}{\textwidth}
		\includegraphics[width=\textwidth]{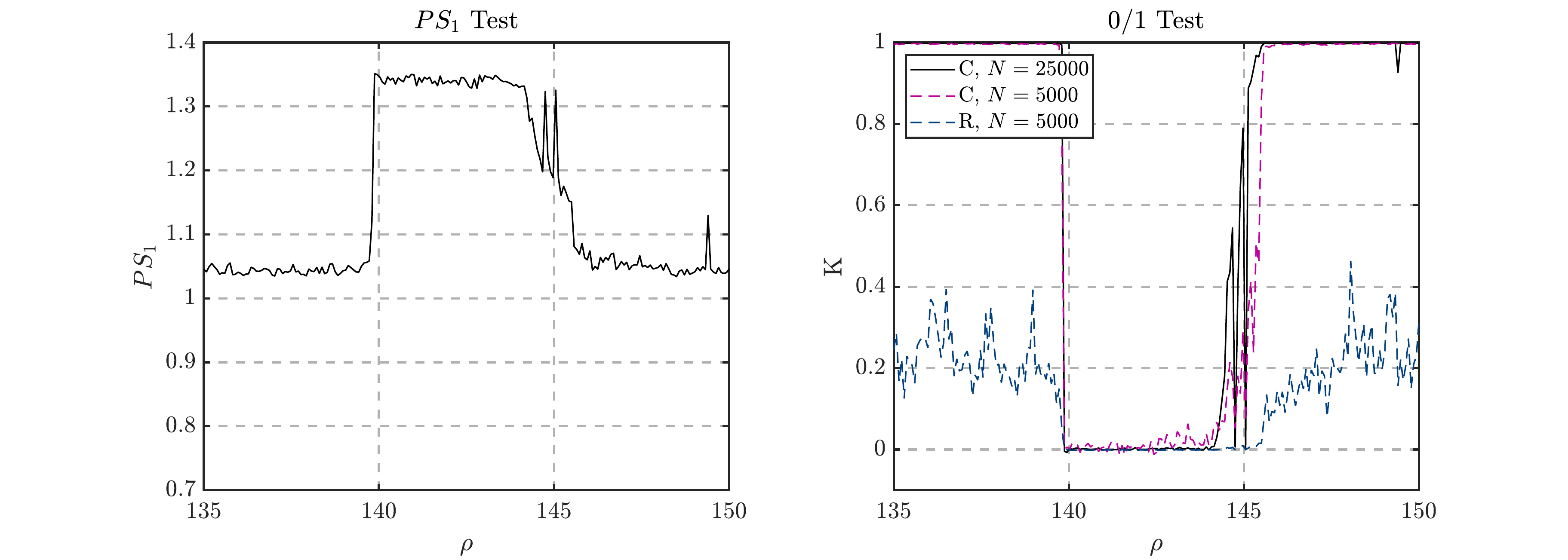}
	\end{subfigure}
	\caption{(top)Bifurcation of the Lorenz model for $\rho\in[135, \ 150]$ and (bottom)The $PS_1$ scores and the 0-1 scores for the Lorenz model for $\rho\in[135, \ 150]$.}
	\label{fig:altscores}
\end{figure}

\textcolor{black}{
	However, it is clear from Figs.~\ref{fig:chaostest_lorenz}, ~\ref{fig:rosslerbif}, and~\ref{fig:logbif} that at times the topological approach does not separate time series data as cleanly as the 0--1 correlation test. 
	Specifically, the transition from chaotic to periodic behavior is at times less obvious in comparison to the original 0--1 test (see Figs.~\ref{fig:rosslerbif} and \ref{fig:logbif}).  
}

\subsubsection{Alternative Parameters of the Lorenz Model}

\textcolor{black}{
	In addition to testing across multiple systems, we also explore the Lorenz model of Section~\ref{lorenz}, but this time with the system parameters $\sigma=10$, $\beta=0.1$, and $\rho\in[135,\ 150]$. \textcolor{black}{Again, the time series are sub-sampled per the minimum significant frequency method described in section~\ref{a_tda_approach} so that $2f_{\rm max}<f_s<4f_{\rm max}$.}
	While the periodic scores for $PS_1$ are consistent with the other models (approximately $1.4$), the $PS_1$ scores for chaotic time series settle near 1.05. 
	On the other hand, while 0-1 correlation test separates this data in a binary fashion, the 0--1 regression test does not.  Figure~\ref{fig:altscores} shows the bifurcations over this span of $\rho$ along with the 0--1 and $PS_1$ scores. Again, the extended time 0--1 correlation score ($N=25000$) is used as a reference score.  }

\textcolor{black}{
	Note that the weak chaos exhibited by the Lorenz model in this parameter range is not to be confused with PPC. An application of the test given in~\cite{Wernecke2017} on the full state-space of the model finds that for all instances of $\rho\in[135, \ 150]$, the cross-correlation $C_{12}$ and scaling exponent $\nu$ either both go to zero for $t>T_\lambda$ or both retain a finite nonzero value. Therefore, the model is either fully chaotic or fully periodic in this range by the definitions in~\cite{Wernecke2017}. 
}

\textcolor{black}{
	With this we conclude that the $PS_1$ scores can be less sensitive to weak chaos than the 0--1 correlation test. Nevertheless, the $PS_1$ scores still clearly identify the {\textit{shift point}} between periodic and chaotic regimes. This decreased sensitivity of the $PS_1$ score is balanced by its increased robustness to noise in periodic time series as we show in Section~\ref{robustness_to_noise}. It is also worth stating that the modified 0--1 regression test (the test prescribed to handle noisy data) performs poorly in this parameter range as well; the $PS_1$ scores are in fact more informative here since they yield consistent results for $\rho\in[135, 140]$ and $\rho\in[145, 150]$ while the $K$ scores of the 0--1 regression test do not. 
}

\section{Robustness to Noise}
\label{robustness_to_noise}

\textcolor{black}{
	All of our results thus far have addressed simulated, noise free time series. 
	However, all signals obtained in the physical world are inherently contaminated with noise to some degree, and it is important to consider the effect of this noise on new time series analysis methods. 
	Thus, we investigate the robustness of the $PS_1$ scores to noise and compare their performance to both the correlation and regression 0--1 methods. 
}
\begin{figure}[b!]
	\centering
	\begin{subfigure}{.5\textwidth}
		\centering
		\includegraphics[width = \textwidth]{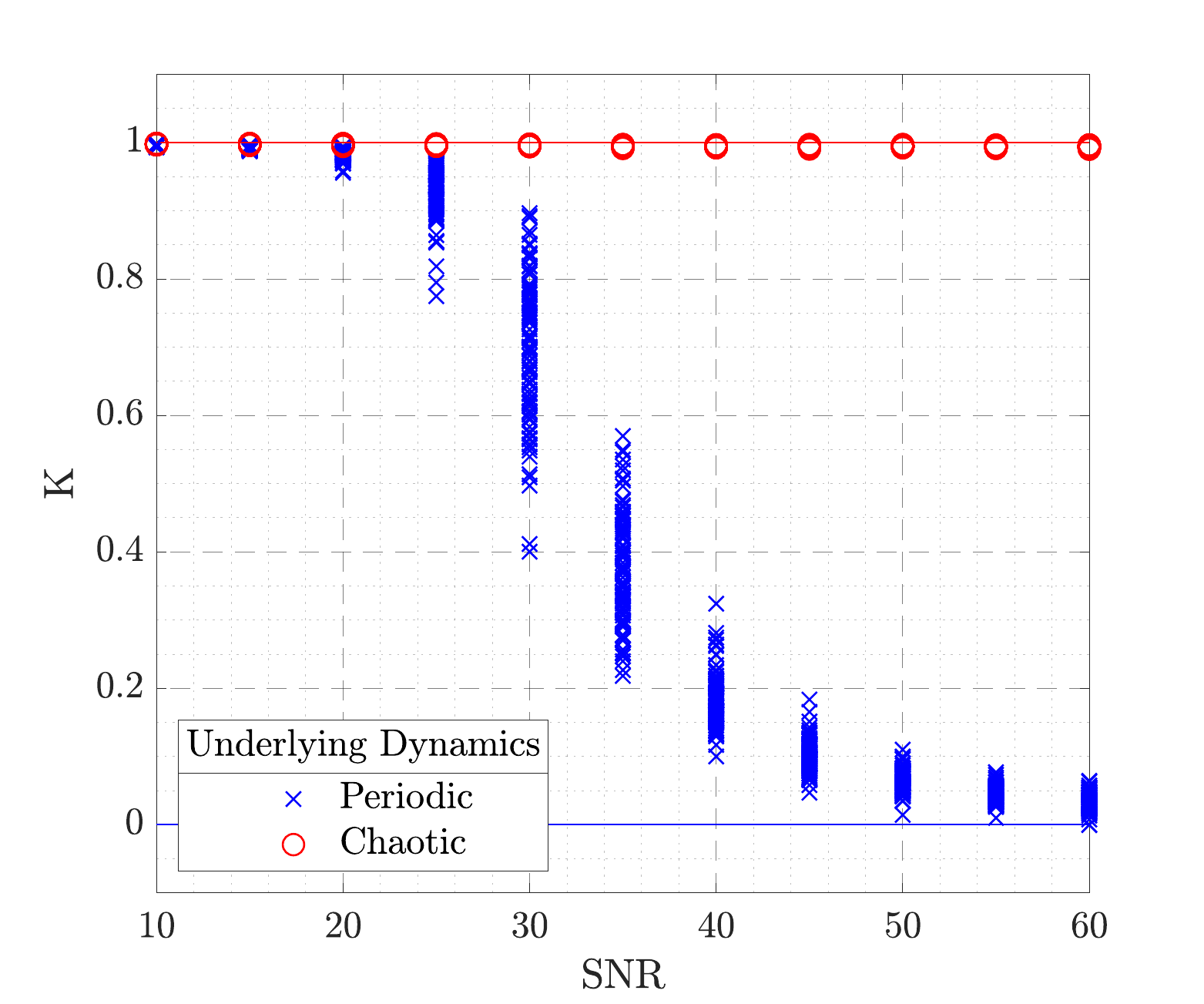}
	\end{subfigure}%
	\begin{subfigure}{.5\textwidth}
		\centering
		\includegraphics[width = \textwidth]{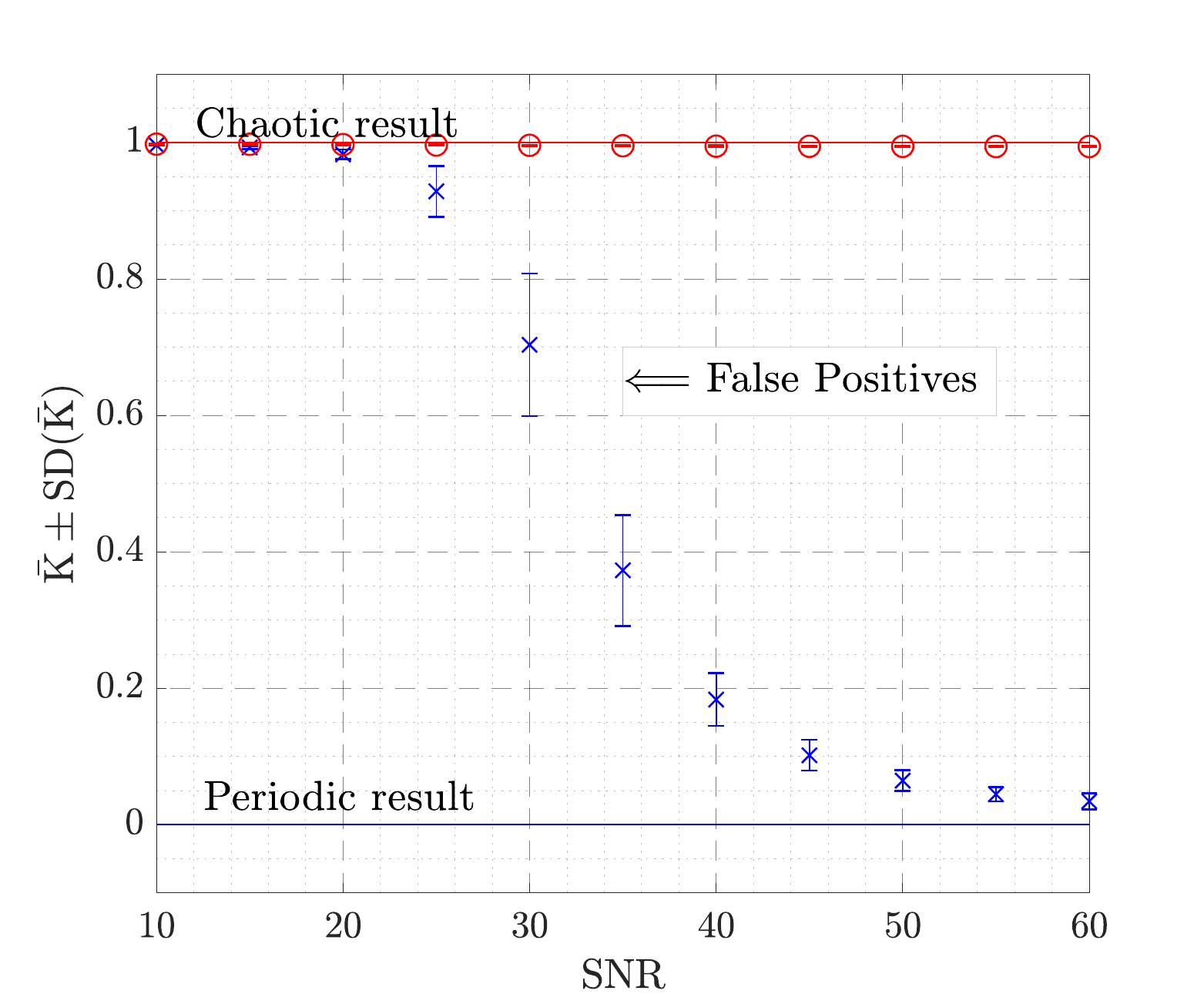}
	\end{subfigure}
	\caption{Effects of  noise on the Correlation 0--1 chaos test with one-hundred tests being performed for each dynamic regime starting from noise levels of SNR= 50dB to SNR=10 dB with all points plotted (left), and (right) a plot of the mean $K$ values as well as error bars given by one standard deviation. It can be seen that 30 dB lead to ambiguous results for periodic time series, and noise levels above 30 dB lead to false positives.}
	\label{fig:GotNoise}
\end{figure}

\begin{figure}[b!]
	\centering
	\begin{subfigure}{.5\textwidth}
		\centering
		\includegraphics[width=\textwidth]{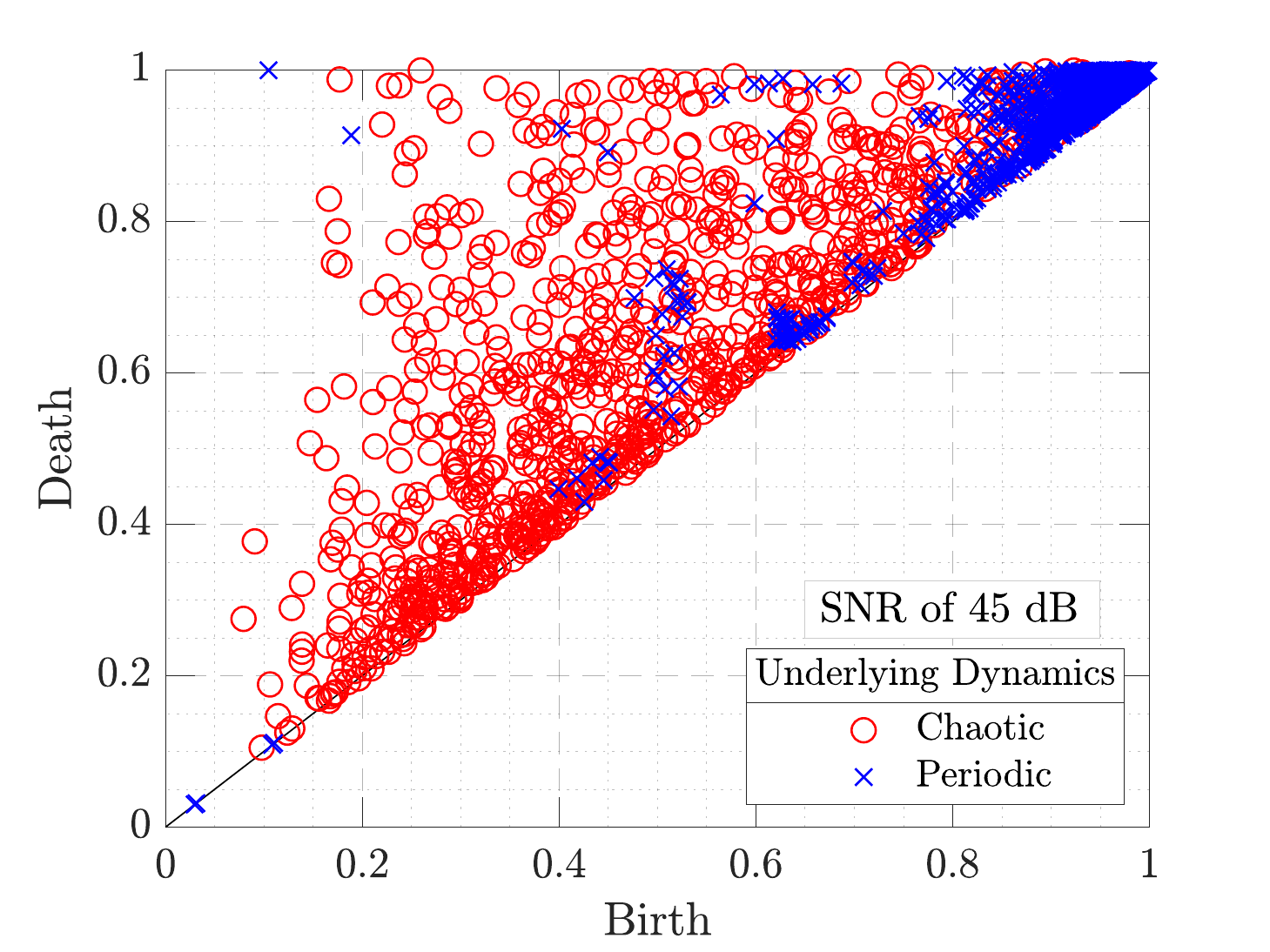}
	\end{subfigure}%
	\begin{subfigure}{.5\textwidth}
		\centering
		\includegraphics[width=\textwidth]{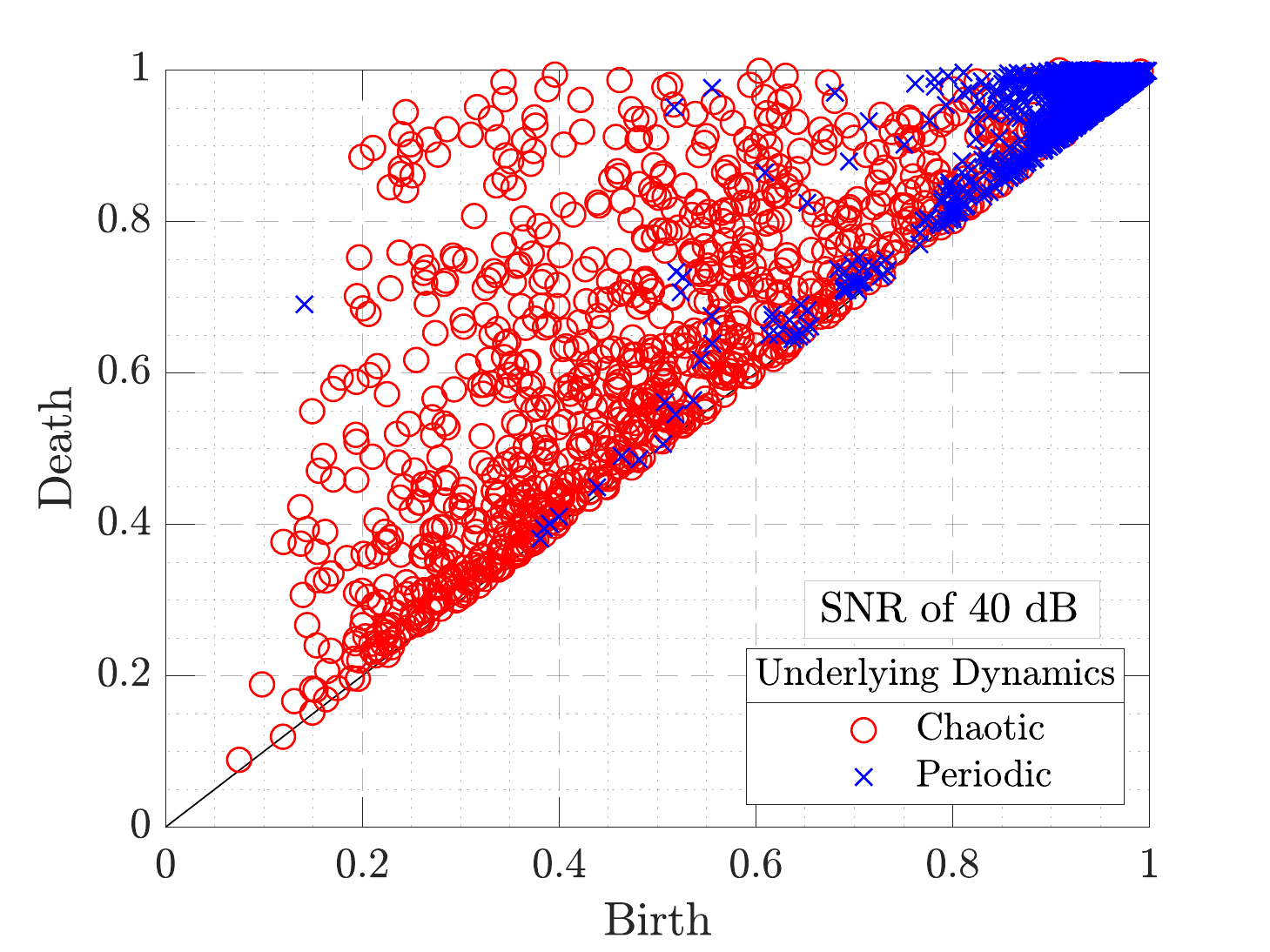}
	\end{subfigure}
	\begin{subfigure}{.5\textwidth}
		\centering
		\includegraphics[width=\textwidth]{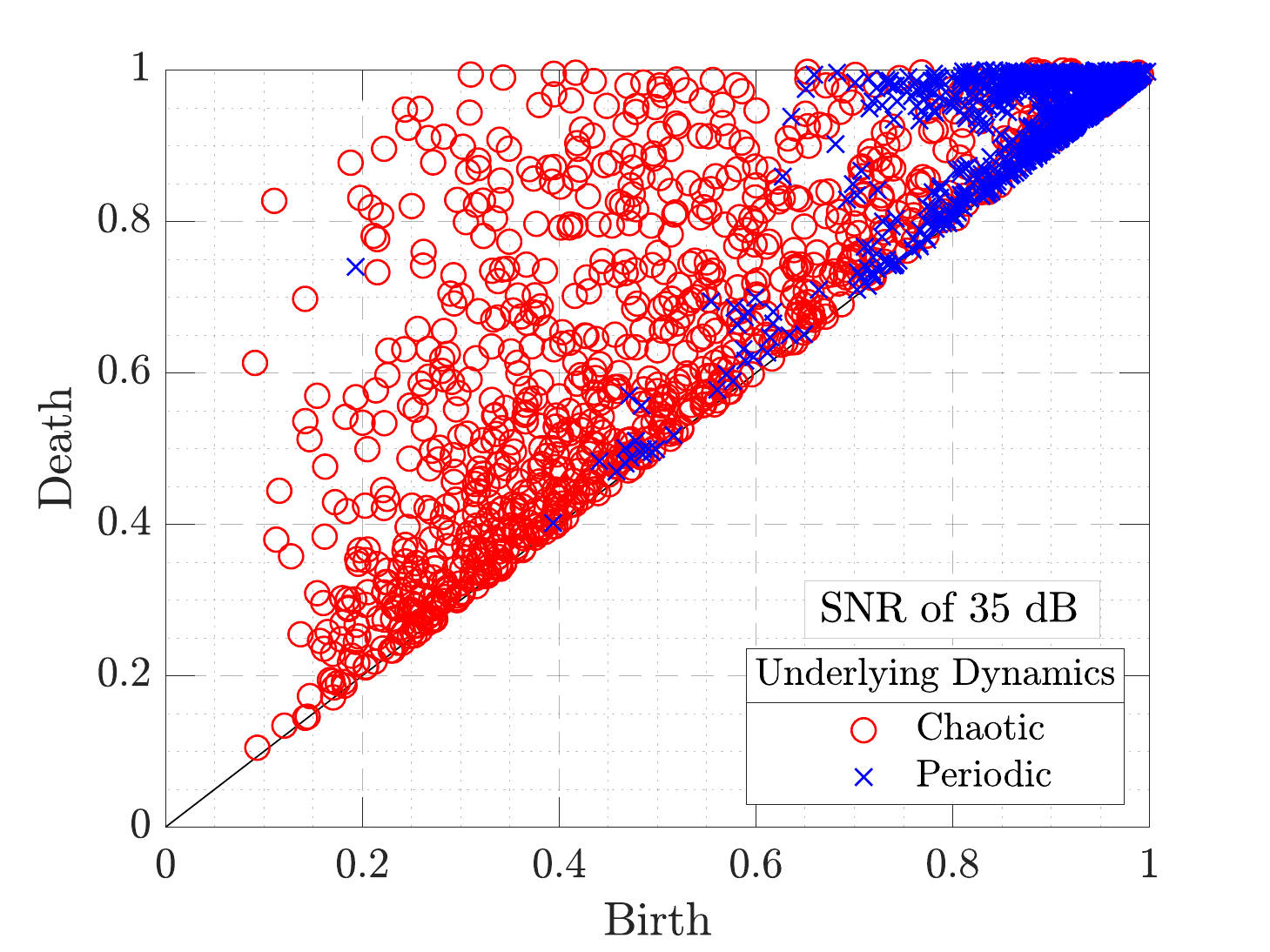}
	\end{subfigure}%
	\begin{subfigure}{.5\textwidth}
		\centering
		\includegraphics[width=\textwidth]{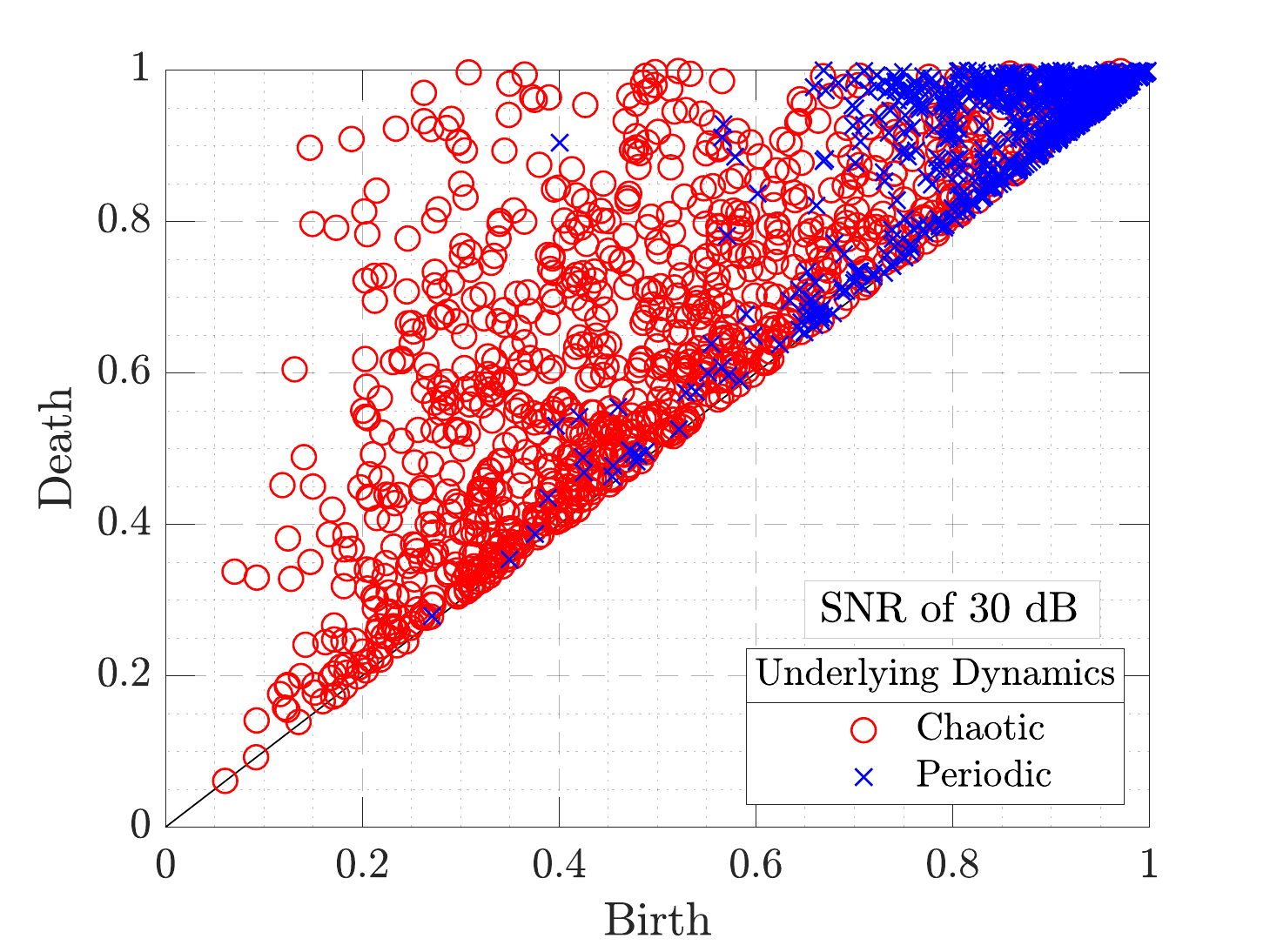}
	\end{subfigure}%
	\caption{The shifts in birth and death time trends as noise is added to the original time series. Shown here is the effect of the noise levels of $\mathrm{SNR}=45$ dB, $\mathrm{SNR}=40$ dB, $\mathrm{SNR}=35$ dB, and  $\mathrm{SNR}=30$ dB. The diagram representing each noise level is composed of multiple realizations of periodic and chaotic time series generated with the Lorenz model for $\sigma =10$, $\beta = 8/3$, and $\rho\in(180.3, \ 181.3)$. The stability of the persistence points with respect to noise levels is clearly illustrated. }
	\label{fig:Pers_Diagrams}
\end{figure}
\textcolor{black}{
	While the correlation 0--1 test does an excellent job in distinguishing chaos from periodicity, the increased sensitivity makes it more susceptible to noise~\cite{Gottwald2009}.  
	A brief noise sensitivity demonstration is given in Fig.~\ref{fig:GotNoise} to exhibit this for simulations of the Lorenz system. This is done by averaging the median $K_c$ values as computed by Eq.~\eqref{eq:corr} for time series containing signal-to-noise ratios (SNRs) ranging from 60 to 10 decibels (dB).
	The average $K$ at each SNR is computed using 100 realizations of these noisy time series.  
	Figure~\ref{fig:GotNoise} shows that low noise levels (i.e. high SNR) do not affect the resulting K score of the 0--1 correlation test. 
	However, at an SNR of 40 dB, the test begins to yield an ambiguous (non-binary) result. 
	Noise levels above an SNR of 30 dB consistently produce a false-positive indicator for periodic time series.  
	These negative effects on the 0--1 test in the presence of noise were discussed in~\cite{Gottwald2005, Gottwald2008}, These references recommend using the 0--1 regression test with noisy time series which was shown to outperform tangent space methods for chaos detection in noisy signals~\cite{Gottwald2005,Gottwald2009}. 
}

\textcolor{black}{
	We start by showing the effect of increasing the level of noise on the structure of the persistence diagrams. 
	Figure~\ref{fig:Pers_Diagrams} shows that these diagrams remain stable as the SNR is varied between 45 and 30 dB. 
	The reason that persistent homology provides this stability is its ability to emphasize the main features of a data set and ignore minor perturbations~\cite{Cohen-Steiner2006}. 
	In other words, the topological structure of the data does not typically change in the presence of moderate noise.
	However, that structure starts to break down once noise levels reach the point at which the salient topological features of the KDE are no longer distinguishable from noise. 
}

\begin{figure}[h!]
	\centering
	\begin{subfigure}{\textwidth}
		\centering
		\includegraphics[width = \textwidth]{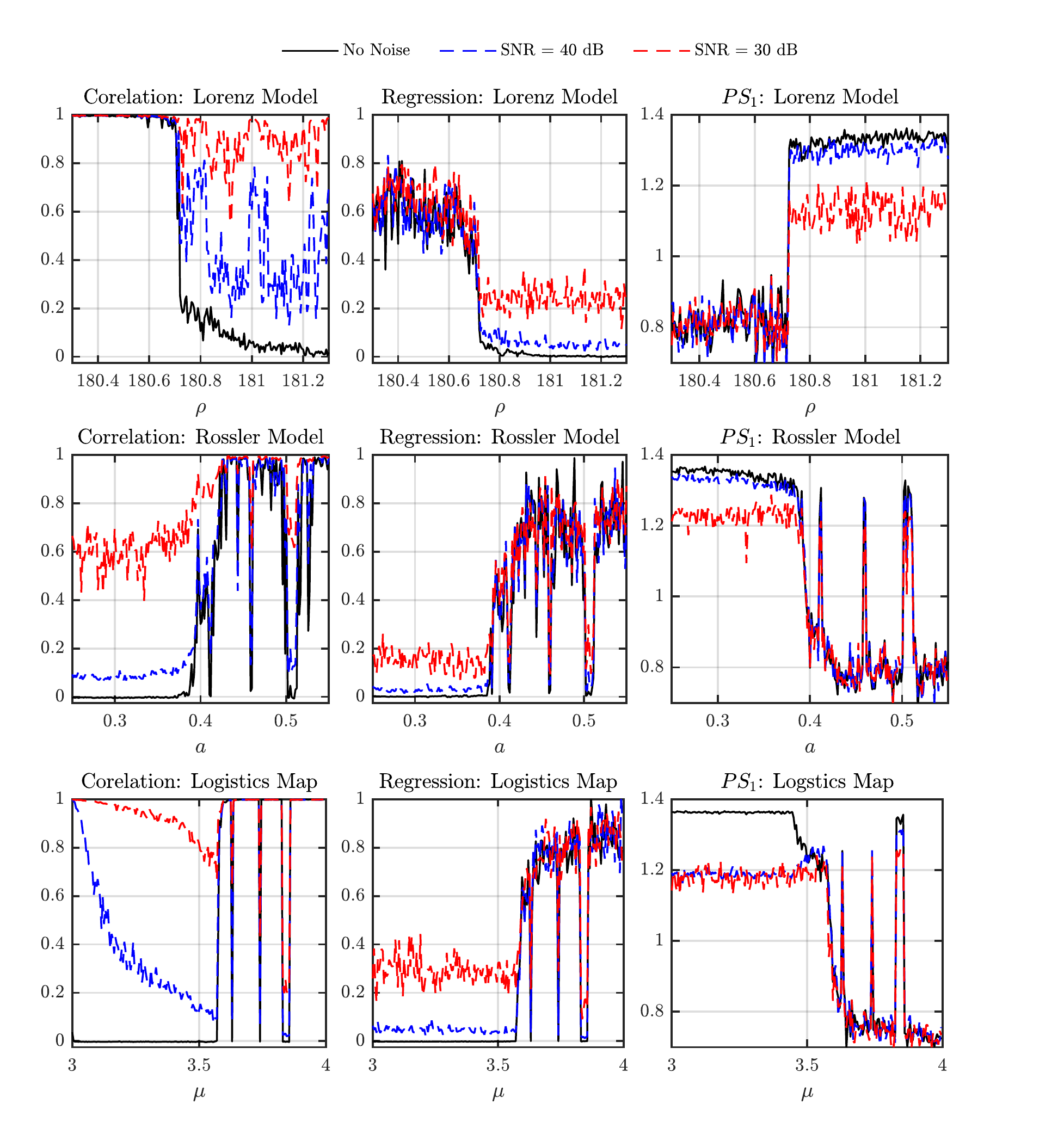}
	\end{subfigure}%
	\caption{(top) The effects of added noise to the (top) Lorenz model, (middle) Rossler model, and (bottom) Logistic map for (left) the 0--1 correlation method (1 or 0, respectively), (middle) to 0--1 modified regression method, and (right) the $PS_1$ scores. The 
		effects of noise are comparable between the modified regression 0--1 test and the $PS_1$ scores, and both of these methods are shown to be far more robust than the 0--1 correlation method.
	}
	\label{fig:noise_comparison}
\end{figure}

\textcolor{black}{
	The robustness of the persistence--based $PS_1$ score to noise is further illustrated with Fig.~\ref{fig:noise_comparison}. 
	Figure~\ref{fig:noise_comparison} shows the effects of noise on the respective detection methods for the Lorenz model, Rossler model, and Logistic map for SNR levels of 40 and 30 dB.  We note that a minimum SNR of 15 dB is often necessary to extract any useful information from a signal.
	The results obtained from the $PS_1$ scores as well as the modified 0--1 test appear stable for 40 dB of noise; however, deviations from the nominal noise-free cases are apparent when the noise levels reach 30 dB. In contrast, the correlation 0--1 test appears to be greatly affected by the noise showing that the trade off of the increased sensitivity to the test is the inability to effectively handle noise.  
}


\textcolor{black}{
	For the Lorenz model, the last column of Fig.~\ref{fig:noise_comparison} shows that the $PS_1$ scores are both stable and binary when noise is added to the signal. 
	While the modified regression method performs well, it does not appear to separate as cleanly as the $PS_1$ scores. 
	Similar results are shown for the Rossler model; the correlation method quickly separates from the nominal case when noise levels increase while the modified regression method and the $PS_1$ scores remain relatively unaffected. 
	For both the Rossler model and the Logistic maps, the 0--1 correlation test indicates false positives for wide regions of periodic parameters while the persistence scores do not deviate dramatically from what is found in the noise-free data.
	The $PS_1$ scores are comparable to the modified regression 0--1 scores when examining the results of the Logistic map as well. An interesting observation to note is that the correlation method yields a false positive for the periodic parameter $\mu$ near 3.0 even for small noise levels. 
}

\subsection{Effect of the Kernel width} 
\label{Kernel}

\begin{figure}[h!]
	\begin{subfigure}{\textwidth}
		\centering
		\includegraphics[width = \textwidth]{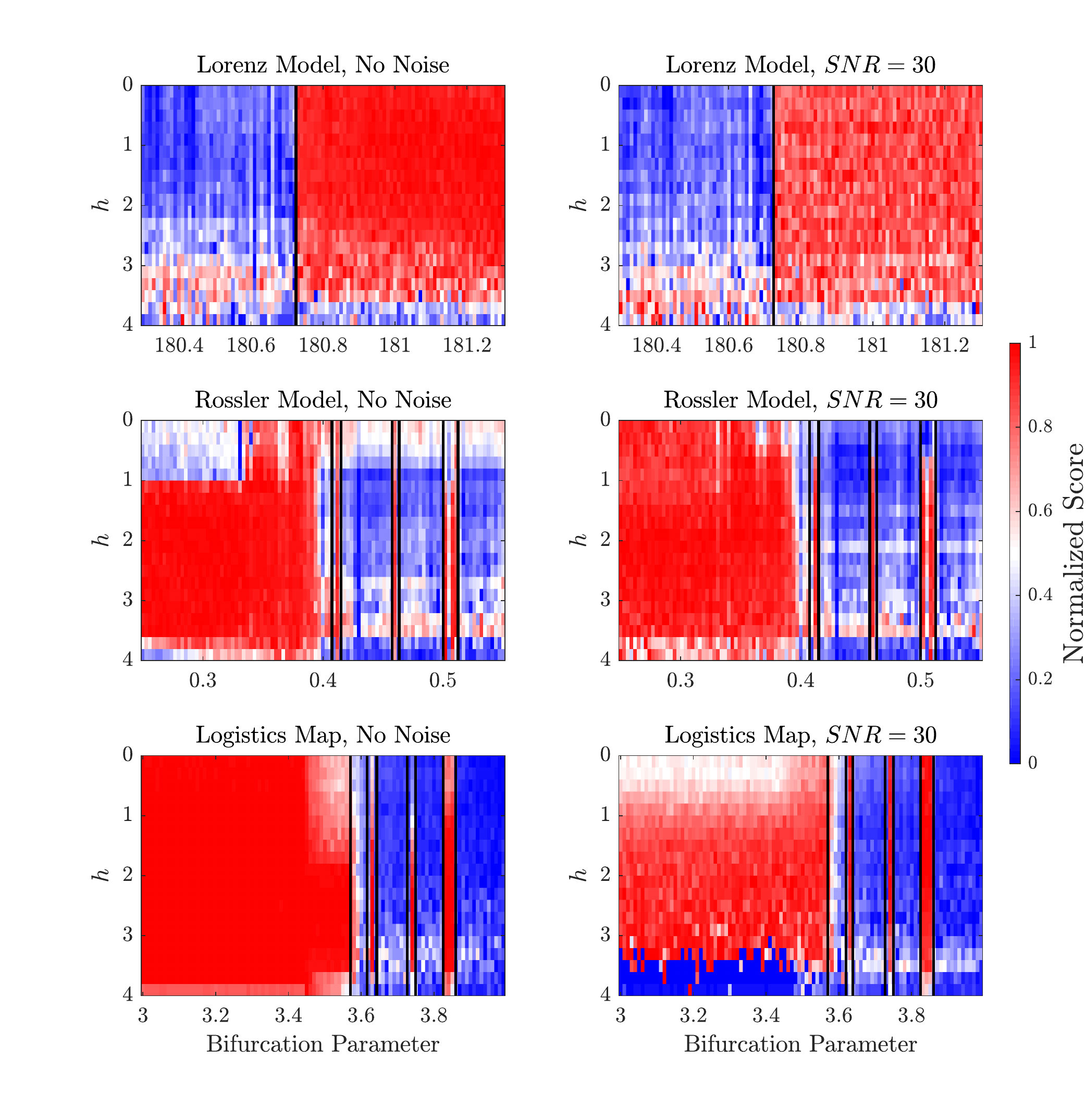}
	\end{subfigure}
	\caption{The normalized persistence scores for (left) no noise and (right) an SNR of 30 dB. The results are given for the Lorenz model, Rossler model, and Logistics map for data with no noise as well as for and SNR of 30 dB. The color values indicate the value of persistence scores at each bifurcation parameter over a range of the kernel width $h\in[0.1, 4]$.}
	\label{fig:kernstudy}
\end{figure}

\textcolor{black}{
	The kernel width $h$ of Eq.~\eqref{eq:Gaussian_filter} plays an important role in the final persistence score. 
	The value of $h=1.3$ is selected based on a parameter study of the systems across their bifurcation parameters whereby 
	the value of $h$ is varied between 0.1 and 4. 
	The resulting persistence scores are plotted to reveal which kernel parameter provides the best qualitative separation between periodic and chaotic parameter ranges for both clean and noisy (SNR = 30 dB) time series.
	Figure~\ref{fig:kernstudy} shows the effect of the $h$ on the persistence scores.
	Note that Fig.~\ref{fig:kernstudy} displays the normalized persistence score $\widetilde{PS}_1$, which is normalized to have a maximum of one and minimum of zero.  
}

For both the Lorenz model and Logistics map, there is much better contrast between $PS_1$ scores generated from periodic time series versus chaotic time series when $h<2.0$. 
For the Rossler model, the cleanest separation of chaotic and periodic regimes occurs at $1<h<3.5$. 
For the Logistics map, the kernel width does not appear to be a factor for the $PS_1$ scores ability to appropriately differentiate the classes when no noise is present. However, when the SNR is 30 dB the best range for separation is $1<h<2.5$. Thus, it is concluded that for the systems explored in this paper any value $1.0 h \leq 2.0$ will sufficiently distinguish between chaos and periodicity for both clean and noisy time series. Note that if a finer grid resolution is chosen for the KDE construction, this optimal bandwidth range is likely to change.

\section{Conclusion}
\label{conclusion}

Gottwald and Melbourne’s 0-1 test for chaos uses a one dimensional time series to drive a two-dimensional system in a $p$-$q$ space. The test is based on showing that when the original system’s dynamics are regular, the resulting $p$-$q$ trajectories are typically bounded whereas chaotic dynamics lead to diffusive trajectories. Gottwald and Melbourne used these observations to define two flavors of the test: the 0-1 correlation test, and the 0-1 modified regression test. 
\textcolor{black}{
	The former is a more sensitive test but it incurs larger errors in the presence of noise, while the latter provides better results in the presence of noise.  However, mathematical justification has been provided for the 0--1 correlation test~\cite{Gottwald2009a} while the same cannot be said for the modified 0--1 regression test. 
}

This work builds upon the results of Gottwald and Melbourne and identifies the dynamics of the system that generated the time series by examining the geometry of the density of the $p$-$q$ trajectories using persistent homology. Specifically, A Kernel Density Estimate (KDE) is obtained using the $p$-$q$ projections and the corresponding 0D sub-level set persistence is computed and summarized in persistence diagrams. Several projections are obtained for the same time series to acquire ensemble of persistence diagrams that encode the shape of the corresponding KDEs. 
Based on observations of 0D persistence on normalized KDEs, we define the persistence score $PS_1 \in [0, \sqrt{2}]$ which represents the mean distance from the origin for the points in the persistence diagram. $PS_1$ assumes values close to $\sqrt{2}$ when the time series is periodic, but its values are below $\sqrt{2}$ for chaotic signals.

The KDEs are smoothed by a Gaussian filter with kernel width $h$. Fig.~\ref{fig:kernstudy} shows the parameter study which leads to the bandwidth selection of $1.0\leq h\leq 2.0$, and it is found that the results of $PS_1$ scores are affected by the kernel bandwidth. 

The 0--1 regression and correlation test as well as the $PS_1$ scores are applied to the PPC dynamics found in the Lorenz model for $\rho\in[180.7,181,1]$. While Fig.~\ref{fig:chaostest_lorenz} shows that each of these tests report PPC data as periodic in the short-time evaluation ($N=5000$), Fig.~\ref{fig:conv} shows that each test fails to converge on a binary results for PPC data in the long-time limit ($N=10^5$).

For noise-free time series, Figs.~\ref{fig:chaostest_lorenz}, \ref{fig:rosslerbif}, and \ref{fig:logbif} shows that our $PS_1$ score successfully delineates between chaotic and periodic behavior albeit not with as clear of a separation as the 0-1 correlation test. 
In contrast, Fig.~\ref{fig:noise_comparison} shows that when noise is added, the $PS_1$ score maintains a consistent performance owing to the stability of persistence diagrams. 
In this case, comparisons to the more noise robust 0-1 modified regression test show that $PS_1$ provides clearer separation especially for high levels of noise, or equivalently, low levels of Signal to Noise Ratio (SNR).
If a periodic parameter is known, a gauge value may be obtained and then the test described in~\cite{Gottwald2016} can be performed. 
Otherwise, either the modified regression test or the $PS_1$ scores should be evaluated. 
The $PS_1$ scores perform just as well the modified regression test in the presence of noise, and the $PS_1$ scores separate data effectively for clean time series.



\section*{Acknowledgment}
The authors would like to acknowledge Dr. Elizabeth Munch for the helpful insight she provided regarding persistent homology and its applications. This material is based upon work supported by the National Science Foundation under Grant Nos. CMMI-1759823 and DMS-1759824 with PI FAK.

\bibliographystyle{ieeetr}
\bibliography{ChaosPaper} 
\end{document}